\documentclass[11pt, twoside, leqno, twocolumn]{article}
\usepackage[english]{babel}
\usepackage{ltexpprt}
\usepackage[utf8]{inputenc}
\usepackage{microtype}
\usepackage{listings}
\newcommand{\anonymize}[1]{#1~}

\let\accentvec\vec

\let\vec\accentvec

\newcommand{\E}{\textnormal{\textrm{E}}}
\usepackage{enumitem}

\newcommand\ips{\textsf{IPS$^4$o}}
\newcommand\blockqs{\textsf{BlockQS}}
\newcommand\stdsort{\textsf{stdsort}}
\newcommand\lomutoone{\textsf{BlockLomuto1}}
\newcommand\lomutotwo{\textsf{BlockLomuto2}}

\usepackage{amsmath}
\usepackage{amssymb}
\usepackage{amsfonts}
\usepackage{verbatim}
\usepackage[textsize=tiny]{todonotes}
\usepackage{algorithm}
\usepackage[noend]{algpseudocode}

\usepackage{pgfplots}
\usetikzlibrary{patterns}
\usepgfplotslibrary{groupplots,colorbrewer}
\pgfplotsset{
  legend style = {font=\ttfamily}
}
\pgfplotsset{
cycle list/Set1-5,
cycle multiindex* list={
mark list*\nextlist
Set1-5\nextlist
},
}

\usepackage{url}

\definecolor{gruen}{rgb}{0.95,0.95,0.95}
\definecolor{dunkelgruen}{rgb}{0.35, 0.7, 0.67}
\definecolor{hellgruen}{rgb}{0.84, 0.7, 0.39}

\title{Simple and Fast BlockQuicksort using Lomuto's Partitioning Scheme%
}

\author{\anonymize{Martin Aumüller} \and \anonymize{Nikolaj Hass}}
\begin{document}
\maketitle
\begin{abstract}
This paper presents simple variants of the BlockQuicksort algorithm described by Edelkamp and Weiss (ESA 2016). The simplification is achieved by using Lomuto's partitioning scheme instead of Hoare's crossing pointer technique to partition the input. To achieve a robust sorting algorithm that works well on many different input types, the paper introduces a novel two-pivot variant of Lomuto's partitioning scheme. A surprisingly simple twist to the generic two-pivot quicksort approach makes the algorithm robust. The paper provides an analysis of the theoretical properties of the proposed algorithms and compares them to their competitors. The analysis shows that Lomuto-based approaches incur a higher average sorting cost than the Hoare-based approach of BlockQuicksort. Moreover, the analysis is particularly useful to reason about pivot choices that suit the two-pivot approach. An extensive experimental study shows that, despite their worse theoretical behavior, the simpler variants perform as well as the original version of BlockQuicksort.
\end{abstract}


\section{Introduction}\label{sec:introduction}
Sorting a sequence of elements in an efficient way is one of Computer Science's fundamental problems. It is one of the most important core routines in many algorithms, and does not seem to lose relevance as we get more and more data to work on every day. 

While it has been more than 55 years since the quicksort algorithm was described by Hoare in~\cite{hoare62}, variants of it are still relevant today.
This especially shows when looking at sorting algorithms implemented in the standard libraries of modern programming languages, such as \textsf{C++}, which uses a worst-case aware quicksort variant called introsort \cite{Musser97}, Oracle's \textsf{Java}, which uses a two-pivot quicksort algorithm~\cite{nebel12}, and Microsoft's \textsf{C\#}, which uses a traditional median-of-three quicksort approach.

In the pursuit of improving the performance of quicksort-based implementations, a long line of papers~\cite{NebelWM15,nebel12,Kushagra14,AumullerDK16,Wild18} have looked at the benefits of introducing more than one pivot in the algorithm. Here, \cite{Kushagra14,AumullerDK16,Wild18} showed that the memory access behavior improves with a small number of pivots --- an effect that cannot be achieved by other more traditional means such as choosing a pivot from a sample of elements~\cite{SedgewickAnalysis}. However, better cache behavior leaves the problem of branch mispredictions unsolved, which is one of the largest bottleneck's in today's processor architectures. 


As shown by Brodal and Moruz in~\cite{BrodalM05}, mispredicting conditional branches, i.e., mispredicting the outcome of the comparison of two elements of the input, are a necessity in the regime of comparison-efficient sorting algorithms. 
This means other tools are needed to reduce the performance penalty that these mispredictions incur, which can be as large as 15 cycles on a modern Intel i7 CPU \cite{HennessyP11}, and even larger on other architectures~\cite{KaligosiS06}. 
As we will describe in Section~\ref{sec:prelims}, under certain circumstances modern processors can avoid branches using so-called conditional move-operations, among others, which were used in a number of implementations of quicksort variants~\cite{SandersW04,ElmasryKS12,EdelkampW16,AxtmannWF017}. More detailed information about these methods is deferred to the related work part of the introduction. 

In the branch-free regime of~\cite{SandersW04,EdelkampW16,AxtmannWF017}, a lot of complexity has been introduced to the algorithm. In fact, Axtmann et al.~\cite{AxtmannWF017} describe code complexity as the main drawback of their approach, stating ``\emph{formal verification of the correctness of the implementation might help to increase trust in the remaining cases.}''~\cite[p. 12]{AxtmannWF017}. 
The BlockQuicksort approach of Edelkamp and Weiss~\cite{EdelkampW16}  has a very elevated code complexity as well, using about 300 lines of code for the partitioning procedure.

 The main contribution of this paper are simple, fast, and robust variants of branch-free quicksort-based algorithms. Thus, introducing more complexity is not necessarily required to improve performance. In more detail, the paper presents simpler variants of the BlockQuicksort algorithm described in~\cite{EdelkampW16}. 
 The difference to BlockQuicksort lies in utilizing a partitioning scheme that Bentley in~\cite{Bentley86} attributed to Nico Lomuto, rather than using the traditional ``crossing pointer'' partitioning scheme by Hoare as in~\cite{EdelkampW16}. The description of the algorithms is provided in Section~\ref{sec:algorithm}. In there, we describe a one-pivot and a two-pivot approach. 
 %
%
 Introducing an additional pivot allows us to make the algorithm robust and to avoid some of the common performance issues of the Lomuto partitioning scheme, such as the presence of many equal elements~\cite{EdelkampW16}. 

Theoretical properties of the proposed algorithms are discussed in Section~\ref{sec:theory}. The result of this analysis allows us, among others, to reason about good pivot choices for the two-pivot variant. Comparing the Lomuto-based algorithms to the Hoare-based approach used in standard BlockQuicksort, we notice a slightly worse comparison count and a slightly worse memory behavior with regard to the ``scanned elements/memory accesses'' cost measure discussed in \cite{Kushagra14,AumullerDK16,Wild18}. This makes the results of the experiments in Section~\ref{sec:evaluation} quite surprising, which show that the proposed variants can compete with BlockQuicksort. In the same section, we speculate about the reasons for this based on actual measurements from the CPU. 


\subsection{Related Work}

\paragraph*{Tuning Quicksort Algorithms} Quicksort implementations (see Section~\ref{sec:prelims} for a description of the quicksort algorithm) usually use insertionsort to sort subproblems that are below a certain size threshold \cite{SedgewickAnalysis}. Moreover, the pivot is usually chosen from a small sample of elements. In practice, it is usually chosen as the median in a sample of three elements, while theory suggests to choose it as the median of $\Theta(\sqrt{n})$ elements to minimize comparisons~\cite{MartinezR01}.  

\paragraph*{Super-Scalar Samplesort}

Sanders and Winkel~\cite{SandersW04} provided an engineered implementation of samplesort, a variant of quicksort in which many pivots are used to distribute the input to buckets. They used conditional moves to traverse the heap-like tree used to find the bucket of an element. The algorithm uses two phases, classification and rearranging elements, and needs extra space in the size of the input.

\paragraph*{Tuned Quicksort}
Elmarsy et al.~\cite{ElmasryKS12} described an engineered quicksort variant that uses a branch-free variant of Lomuto's partitioning scheme using conditional moves.

\paragraph*{BlockQuicksort}
 BlockQuicksort as introduced by Edelkamp and Weiss in~\cite{EdelkampW16} generalizes Hoare's partitioning scheme using blocks to handle misplaced elements. Filling these two blocks, one block for each of the two pointers used in Hoare's partitioning scheme, with references to misplaced elements is done branch-free using conditional moves. When these blocks are near-full, elements are moved into a correct position with regard to the pivot choice. When the two pointers are close at the end of the partitioning phase, the code has to handle many edges cases to produce a correct partition. To this end, BlockQuicksort uses a ``clean-up phase'' when partitioning is done.


\paragraph*{Multi-Pivot Quicksort}
\cite{hennequin,NebelWM15,nebel12,Kushagra14,Aumuller15,AumullerD16,AumullerDK16,Wild18} studied benefits of using a small number of pivots in quicksort.  This line of research showed that this approach improves the memory access pattern drastically~\cite{AumullerDK16,Wild18}, an improvement that cannot be achieved by other means such as choosing a good pivot by sampling input elements. In short, multi-pivot quicksort allows to make fewer accesses to array cells to sort the input, an
effect that translates into better running times. The most popular multi-pivot algorithm is the two-pivot \textsf{YBB} algorithm~\cite{nebel12} used in Oracle's \textsf{Java} since version 7. 

\paragraph*{\ips} \cite{AxtmannWF017} provides an in-place samplesort variant of \cite{SandersW04} by combining it with the idea of using blocks from BlockQuicksort. Instead of two blocks, they use a large number of blocks, one for each bucket. These blocks are filled in a branch-free way using the classification strategy from \cite{SandersW04}. All elements are moved  in blocks and a final rearrangement step is needed to swap blocks to correct positions. Again, a lot of care is needed in the clean-up phase. \ips~lends itself to parallelization, as demonstrated by the experiments in \cite{AxtmannWF017}. 

\subsection{Our Contributions}

This paper contains the following contributions:
\begin{itemize}[noitemsep, nolistsep]
    \item We present variants of BlockQuicksort using Lomuto's partitioning scheme (Algorithm~\ref{algo:single:pivot:partitioning}, Algorithm~\ref{algo:dual:pivot:partitioning}) and evaluate them experimentally. Comparing the implementations, the proposed variants save lines of code by a factor 5 to 8 times compared to \cite{EdelkampW16} and are easier to understand.
    \item We show how using a simple twist in two-pivot quicksort makes the algorithm more robust with respect to different input types.
    \item We analyze theoretical properties of our variants, which among others allows us
        to reason about good pivot choices. 
\end{itemize}
We believe the algorithms to be so simple to be described in a textbook on Algorithm Engineering.
\section{Preliminaries}\label{sec:prelims}

\paragraph*{Outline of a One-Pivot Quicksort Algorithm}

We assume that the input to be sorted resides in an array $A[1..n]$. Classical quicksort (Algorithm~\ref{algo:single:pivot:qs}) sorts the array $A$ as follows. If $n \leq 1$, do nothing. Otherwise, choose an element $p$ from $A$ as pivot. Next, \emph{partition} the input such that $p$ resides in $A[i]$, $A[1..i - 1]$ contains elements smaller than $p$, and $A[i + 1..n]$ contains elements at least as large as $p$. Then, sort $A[1..i - 1]$ and $A[i + 1..n]$ recursively.

\begin{algorithm}[t!]
    \caption{One-Pivot-Quicksort}\samepage\label{algo:single:pivot:qs}
    \textbf{procedure} \textsf{OnePivotQuicksort}($\textit{A}[1..\textit{n}]$)
    \begin{algorithmic}[1]
        \If{$\textit{n} > 1$}
            \State $\text{choosePivot}(\textit{A}[1..\textit{n}])$
            \Comment{Pivot resides in $\textit{A}[\textit{n}]$}
            \State $\texttt{i} \gets \text{partition}(\textit{A}[1..\textit{n}])$
            \State $\textsf{OnePivotQuicksort}(\textit{A}[1..\texttt{i} - 1])$
            \State $\textsf{OnePivotQuicksort}(\textit{A}[\texttt{i} + 1..\textit{n}])$
        \EndIf
    \end{algorithmic}
\end{algorithm}

\paragraph*{Lomuto's Partitioning Scheme}

Traditionally, one uses the crossing pointer technique described by Hoare~\cite{hoare62} to partition the input. A simpler partitioning method communicated to Bentley by Lomuto\footnote{To quote \cite[Page~110]{Bentley86}: \emph{A reader of a preliminary draft [of \cite{Bentley86}] complained that the standard two-index method is in fact simpler than Lomuto's, and sketched some code to make his point: I stopped looking after I found two bugs.}} is given as Algorithm~\ref{algo:lomuto}. 

\begin{figure}[t!]
\centering
\begin{tikzpicture}[xscale = 1.05]
\draw (0,0) rectangle (8, 0.5);
\draw[fill=hellgruen!30] (0,0) rectangle (2, 0.5);
\draw[fill=dunkelgruen!30] (2,0) rectangle (5, 0.5);
\draw[fill=dunkelgruen!30] (7.5,0) rectangle (8, 0.5);
\fill[fill=hellgruen!30] (5,0) rectangle (7.5, 0.25);
\fill[fill=dunkelgruen!30] (5,0.25) rectangle (7.5, 0.5);
\draw (7.5, 0) -- (7.5, 0.5);
\draw (2, 0) -- (2, 0.5);
\draw (5, 0) -- (5, 0.5);
\node at (7.75, 0.25) {$p$};
\node at (1, 0.25) {$< p$};
\node at (3.5, 0.25) {$\geq p$};
\node at (6.25, 0.25) {? $\cdots$ ?};
\node at (2.25, -0.25) {\small$\uparrow$};
\node at (2.25, -0.65) {$\texttt{i}$};
\node at (5.25, -0.25) {\small$\uparrow$};
\node at (5.25, -0.65) {$\texttt{j}$};
\end{tikzpicture}
\vspace*{-2.5em}
\caption{Lomuto invariant: $A[1..\texttt{i} - 1]$ consists of elements smaller than $p$, 
$A[\texttt{i}..\texttt{j}-1]$ consists of elements at least as large as $p$; $A[\texttt{j}..\textit{n}-1]$ has not been looked at, which is depicted by filling this part of the array with both colors.} 
\label{fig:lomuto:invariant}
\end{figure}
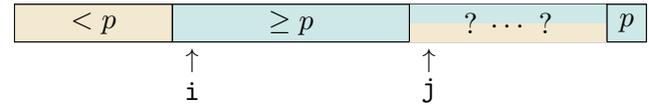

Lomuto's partitioning scheme uses two additional variables $\texttt{i}$ and $\texttt{j}$ and maintains the invariant displayed in Figure~\ref{fig:lomuto:invariant}. The variable $\texttt{j}$ is incremented from $1$ to $n - 1$ using a \textbf{for} loop. When $\textit{A}[\texttt{j}]$ is inspected, it is compared to the pivot $\texttt{p}$. If it is smaller than the pivot, $\textit{A}[\texttt{i}]$ and $\textit{A}[\texttt{j}]$ are swapped, and $\texttt{i}$ is incremented. 

As pointed out in~\cite{Bentley86,Wild16}, Lomuto's partitioning scheme is ``not as fast as Hoare's version''. Theoretically speaking, this is because on average, where the average is taken over all $n!$ permutations of the set $\{1, \ldots, n\}$, it makes three times more swaps than Hoare's partitioning scheme~\cite{LomutoSwaps} and ``scans'' 50\% more elements~\cite{Wild16}.

\begin{algorithm}[t!]
    \caption{Lomuto Partitioning Scheme}\samepage\label{algo:lomuto}
    \textbf{procedure} \textsf{LomutoPartition}($\textit{A}[1..\textit{n}]$)
    \begin{algorithmic}[1]
        \State $\texttt{p} \gets \textit{A}[\textit{n}]$
        \State $\texttt{i} \gets 1$
        \For{$\texttt{j} \gets 1; \texttt{j} < \textit{n}; \texttt{j} \gets \texttt{j} + 1$}
            \If{$\textit{A}[\texttt{j}] < \texttt{p}$}
                \State Swap $\textit{A}[\texttt{i}]$ and $\textit{A}[\texttt{j}]$
                \State $\texttt{i} \gets \texttt{i} + 1$
            \EndIf
        \EndFor
        \State Swap $\textit{A}[\texttt{i}]$ and $\textit{A}[\textit{n}]$\\
        \Return $\texttt{i}$
    \end{algorithmic}
\end{algorithm}

\paragraph*{Outline of a Two-Pivot Quicksort Algorithm}

Two-pivot quicksort (Algorithm~\ref{algo:dual:pivot:qs}) sorts the array $A[1..n]$ as follows. If $n \leq 1$, do nothing. Otherwise, choose two elements $p,q$ with $p \leq q$ from $A$ as pivots. Next, partition the input such that $p$ resides in position $i$ and $q$ resides in position $j$, $A[1..i - 1]$ contains elements smaller than $p$, $A[i + 1 .. j - 1]$ contains elements $x$ with $p \leq x \leq q$ and $A[j + 1..n]$ contains elements larger than $q$. Then, sort $A[1..i - 1]$, $A[i + 1..j - 1]$, and $A[j + 1..n]$ recursively.

\begin{algorithm}[t!]
    \caption{Two-Pivot-Quicksort}\samepage\label{algo:dual:pivot:qs}
    \textbf{procedure} \textsf{TwoPivotQuicksort}($\textit{A}[1..\textit{n}]$)
    \begin{algorithmic}[1]
        \If{$\textit{n} > 1$}
            \State $\text{choosePivot}(\textit{A}[1..\textit{n}])$
            \Comment{Pivots $\textit{A}[1] \leq \textit{A}[\textit{n}]$}
            \State $(\texttt{i}, \texttt{j}) \gets \text{partition}(\textit{A}[1..\textit{n}])$
            \State $\textsf{TwoPivotQuicksort}(\textit{A}[1..\texttt{i} {-} 1])$
            \State $\textsf{TwoPivotQuicksort}(\textit{A}[\texttt{i} {+} 1..\texttt{j} {-} 1])$
            \!\!\!\Comment{Sec. \ref{sec:equal:elements}}
            \State $\textsf{TwoPivotQuicksort}(\textit{A}[\texttt{j} {+} 1..\textit{n}])$
        \EndIf
    \end{algorithmic}
\end{algorithm}

Many different partitioning methods for two-pivot quicksort exist, see, for example~\cite{nebel12,AumullerD16}.

\paragraph*{Avoiding Branch Mispredictions}

Today, most CPUs use instruction-level parallelism. As described in \cite[Section~3.13]{HennessyP11}, the biggest problems in exploiting parallelism come from mispredicted branches or cache misses. Branch mispredictions may occur when the code contains conditional jumps, such as \emph{if} statements. When reaching a branch, the CPU decides which branch it follows based on a branch predictor and loads the instructions following the branch into its pipeline. 
When the direction of the branch is mispredicted, the pipeline has to be flushed and the other direction has to be executed. 

In comparison-based sorting algorithms, most branches occur when input elements are compared to each other. Following~\cite{EdelkampW16}, on modern hardware these branches can be avoided as follows in C++:
\begin{itemize}[noitemsep, nolistsep]
    \item By using conditional moves (\texttt{CMOVcc}), which have the form 
        \texttt{i = (x < y) {?} j {:} i;}
    \item By casting a boolean to an integer (\texttt{SETcc}), which 
    has the form
        \texttt{int i = (x < y);}
\end{itemize} 


\section{Lomuto BlockQuicksort}\label{sec:algorithm}


\subsection{One Pivot}

\begin{algorithm}[t!]
    \caption{One-Pivot Block Partitioning}\samepage\label{algo:single:pivot:partitioning}
    \textbf{procedure} \lomutoone($\textit{A}[1..\textit{n}]$)
    \begin{algorithmic}[1]
        \Require $\textit{n} > 1, \text{Pivot in $\textit{A}[\textit{n}]$}$
        \State $\texttt{p} \gets \textit{A}[\textit{n}]$; 
    \State \textbf{integer} $\text{block}[0, \ldots, \textit{B} - 1]$, $\texttt{i}, \texttt{j} \gets 1$, $\texttt{num} \gets 0$
    \While{$\texttt{j} < \textit{n}$}
    \State $\texttt{t} \gets \text{min}(\textit{B}, \textit{n} - \texttt{j} )$;
\For{$\texttt{c} \gets 0; \texttt{c} < \texttt{t}; \texttt{c} \gets \texttt{c} + 1$} 
                \State $\text{block}[\texttt{num}] \gets \texttt{c}$;
            \State $\texttt{num} \gets \texttt{num} +  (\texttt{p} > \textit{A}[\texttt{j} + \texttt{c}])$;
            \EndFor
            \For{$\texttt{c} \gets 0; \texttt{c} < \texttt{num}; \texttt{c} \gets \texttt{c} + 1$} 
        \State \text{Swap $\textit{A}[\texttt{i}]$ and $\textit{A}[\texttt{j} + \text{block}[\texttt{c}]]$}
        \State $\texttt{i} \gets \texttt{i} + 1$
            \EndFor
            \State $\texttt{num} \gets 0$;
        \State $\texttt{j} \gets \texttt{j} + \texttt{t}$;
        \EndWhile
        \State Swap $\textit{A}[\texttt{i}]$ and $\textit{A}[\textit{n}]$;\\
    \Return $\texttt{i}$;
    \end{algorithmic}
\end{algorithm}
\begin{figure}[t!]
\vspace*{-1.5em}
\centering
\begin{tikzpicture}[xscale = 1.0]
\draw[fill=hellgruen!30] (0,0) rectangle (1.5, 0.5);
\draw[fill=dunkelgruen!30] (1.5,0) rectangle (3.5, 0.5);
\draw[fill=dunkelgruen!30] (7.5,0) rectangle (8, 0.5);
\fill[fill=dunkelgruen!30] (4.5,0.25) rectangle (7.5, 0.5);
\fill[fill=hellgruen!30] (4.5,0) rectangle (7.5, 0.25);
\fill[fill=dunkelgruen!30] (3.5,0) rectangle (3.75, 0.5);
\fill[fill=hellgruen!30] (3.75,0) rectangle (4.0, 0.5);
\fill[fill=hellgruen!30] (4,0) rectangle (4.25, 0.5);
\fill[fill=dunkelgruen!30] (4.25,0) rectangle (4.5, 0.5);
\draw (0,0) rectangle (8, 0.5);
\draw (7.5, 0) -- (7.5, 0.5);
\draw (1.5, 0) -- (1.5, 0.5);
\draw (3.5, 0) -- (3.5, 0.5);
\draw[dotted, thick] (5.5, 0) -- (5.5, 0.5);
\node at (7.75, 0.25) {$p$};
\node at (0.75, 0.25) {$< p$};
\node at (2.5, 0.25) {$\geq p$};
\node at (6.5, 0.25) {? $\cdots$ ?};
\node at (1.65, -0.25) {\small$\uparrow$};
\node at (1.65, -0.65) {$\texttt{i}$};
\node at (3.65, -0.25) {\small$\uparrow$};
\node at (3.65, -0.65) {$\texttt{j}$};
\node at (4.65, -0.25) {\small$\uparrow$};
\node at (4.65, -0.65) {$\texttt{j} + \texttt{c}$};
\draw [decorate, decoration={brace,amplitude=5pt},yshift=2pt](3.5,0.5)--(5.5,0.5) node [midway,yshift=15pt]{$B$};
\draw (6.25,-1) rectangle (8.25, -0.5);
\draw (6.25, -1) -- (6.25, -0.5);
\draw (6.75, -1) -- (6.75, -0.5);
\draw (7.25, -1) -- (7.25, -0.5);
\draw (7.75, -1) -- (7.75, -0.5);
\draw (8.25, -1) -- (8.25, -0.5);
\node at (6.5, -0.75) {1};
\node at (7, -0.75) {2};
\node at (7.5, -0.75) {3};
\node at (7.5, -1.25) {\small$\uparrow$};
\node at (7.5, -1.6) {$\texttt{num}$};
\node at (6.25, -1.3) {$\texttt{block}$};

\end{tikzpicture}
\vspace{-1.5em}
\caption{BlockQuicksort with Lomuto's partitioning scheme (Algorithm~\ref{algo:single:pivot:partitioning}). Picture depicts the situation where Algorithm~\ref{algo:single:pivot:partitioning} is currently on Line~7 with $\texttt{c} = 4$. So far, the block contains the indexes 1 and 2, representing that $A[\texttt{j} + 1]$ and $A[\texttt{j} + 2]$ are smaller than $p$. In general, given that $\texttt{c}$ has value $c$ and $\texttt{num}$ is \textit{num}, $\texttt{block}[0..\textit{num} - 1]$ contains the indexes (relative to $\texttt{j}$) of all misplaced elements in $A[\texttt{j}..\texttt{j} + c]$. Unlabeled elements have only half the width of the array cell depicting $p$ in the picture.}
\label{fig:lomuto:block:invariant}
\end{figure}
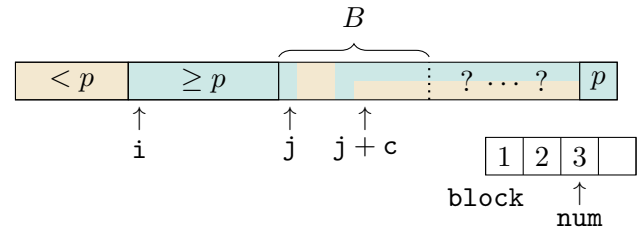

Algorithm~\ref{algo:single:pivot:partitioning} (\lomutoone) describes the full partitioning method that can be plugged into the general quicksort procedure described as Algorithm~\ref{algo:single:pivot:qs}. See Figure~\ref{fig:lomuto:block:invariant} to see the invariant that is kept by the partitioning method. 

Algorithm~\ref{algo:single:pivot:partitioning} is the straight-forward generalization of the standard Lomuto partitioning scheme discussed in the previous section. 
In addition to the pivot $p$ and the  two indexes $\texttt{i}$ and $\texttt{j}$, the algorithm uses an array $\texttt{block}$ that can store $B$ indexes, and a variable $\texttt{num}$. 
Except when there are less than $B$ elements left to consider (see Line~4), the algorithm considers $B$ elements of the input at a time in the \textbf{while} loop that starts on Line~3. 
First, in Lines~5--7, it fills the \texttt{block} array with the indexes of the elements that are smaller than the pivot, i.e., the elements that are misplaced. This is done in a branch-free way by the cast to an integer. Next, in Lines~8-10, all misplaced elements are moved to a final position in the array, \texttt{num} is reset to 0, and $\texttt{j}$ is advanced by one block size. After the loop ends, the pivot is put into place (Line~13) and its position is returned.

We stress that Algorithm~\ref{algo:single:pivot:partitioning} is the full algorithm as used in the experiments. It is easy to describe and, as we will see, performs as well as the BlockQuicksort variant using Hoare's partitioning scheme described in~\cite{EdelkampW16}. The reader is invited to compare \lomutoone~to their \cite[Algorithm~3]{EdelkampW16}, which omits the complicated rearrangement phase. 

\begin{algorithm}[t!]
    \caption{Two-Pivot Block Partitioning}\samepage\label{algo:dual:pivot:partitioning}
    \textbf{procedure} \lomutotwo($\textit{A}[1..\textit{n}]$)
    \begin{algorithmic}[1]
        \Require $n > 1$, \text{Pivots in $\textit{A}[1] \leq \textit{A}[\textit{n}]$}
        \State $\texttt{p} \gets \textit{A}[1]$; $\texttt{q} \gets \textit{A}[n]$; 
    \State \textbf{integer} $\text{block}[0, \ldots, \textit{B} - 1]$
    \State  $\texttt{i}, \texttt{j}, \texttt{k} \gets 2$, $\texttt{num}_{< \texttt{p}}, \texttt{num}_{\leq \texttt{q}} \gets 0$
    \While{$\texttt{k} < \textit{n}$}
    \State $\texttt{t} \gets \text{min}(\textit{B}, \textit{n} - \texttt{k} )$;
\For{$\texttt{c} \gets 0; \texttt{c} < \texttt{t}; \texttt{c} \gets \texttt{c} + 1$} 
            \State $\text{block}[\texttt{num}_{\leq \texttt{q}}] \gets \texttt{c}$;
        \State $\texttt{num}_{\leq \texttt{q}} \gets \texttt{num}_{\leq \texttt{q}} +  (\texttt{q} \geq \textit{A}[\texttt{k} + \texttt{c}])$;
            \EndFor
            \For{$\texttt{c} \gets 0; \texttt{c} < \texttt{num}_{\leq \texttt{q}}; \texttt{c} \gets \texttt{c} + 1$} 
        \State \text{Swap $\textit{A}[\texttt{j} + \texttt{c}]$ and $\textit{A}[\texttt{k} + \text{block}[\texttt{c}]]$}
            \EndFor
            \State $\texttt{k} \gets \texttt{k} + \texttt{t}$;
            \For{$\texttt{c} \gets 0; \texttt{c} < \texttt{num}_{\leq \texttt{q}}; \texttt{c} \gets \texttt{c} + 1$} 
        \State $\text{block}[\texttt{num}_{< \texttt{p}}] \gets \texttt{c}$;
    \State $\texttt{num}_{< \texttt{p}} \gets \texttt{num}_{< \texttt{p}} +  (\texttt{p} > \textit{A}[\texttt{j} + \texttt{c}])$;
            \EndFor
            \For{$\texttt{c} \gets 0; \texttt{c} < \texttt{num}_{< \texttt{p}}; \texttt{c} \gets \texttt{c} + 1$} 
        \State \text{Swap $\textit{A}[\texttt{i}]$ and $\textit{A}[\texttt{j} + \text{block}[\texttt{c}]]$}
        \State $\texttt{i} \gets \texttt{i} + 1$
            \EndFor
            \State $\texttt{j} \gets \texttt{j} + \texttt{num}_{\leq \texttt{q}}$;
    \State $\texttt{num}_{< \texttt{p}}, \texttt{num}_{\leq \texttt{q}} \gets 0$;
        \EndWhile
        \State Swap $\textit{A}[\texttt{i} - 1]$ and $\textit{A}[1]$;
    \State Swap $\textit{A}[\texttt{j}]$ and $\textit{A}[\textit{n}]$;\\
\Return $(\texttt{i} - 1,\texttt{j})$;
    \end{algorithmic}
\end{algorithm}
\begin{figure}[t!]
\vspace*{-1em}
\centering
\scalebox{.92}{
\begin{tikzpicture}[xscale = 1.05, xshift=0.5cm]
\draw[fill=gruen!30] (-0.5, 0) rectangle (0, 0.5);
\draw[fill=hellgruen!30] (0,0) rectangle (1, 0.5);
\draw[fill=gruen!30] (1,0) rectangle (3, 0.5);
\draw[fill=dunkelgruen!30] (3, 0) rectangle (4, 0.5);
\draw[fill=gruen!30] (7.5,0) rectangle (8, 0.5);
\fill[fill=dunkelgruen!30] (4,0) rectangle (4.25, 0.5);
\fill[fill=gruen!30] (4.25,0.25) rectangle (4.75, 0.5);
\fill[fill=hellgruen!30] (4.25,0) rectangle (4.75, 0.25);
\fill[fill=hellgruen!30] (4.75,0) rectangle (7.5, 0.16);
\fill[fill=gruen!30] (4.75,0.16) rectangle (7.5, 0.33);
\fill[fill=dunkelgruen!30] (4.75,0.33) rectangle (7.5, 0.5);

\draw (0,0) rectangle (8, 0.5);
\draw (7.5, 0) -- (7.5, 0.5);
\draw[dotted, thick] (5.5, 0) -- (5.5, 0.5);
\node at (-0.25, 0.25) {$p$};
\node at (7.75, 0.25) {$q$};
\node at (0.5, 0.25) {$< p$};
\node at (2, 0.25) {$p \leq \dots \leq q$};
\node at (3.5, 0.25) {$> q$};
\node at (6.5, 0.25) {? $\cdots$ ?};
\node at (1.05, -0.25) {\small$\uparrow$};
\node at (1.05, -0.65) {$\texttt{i}$};
\node at (3.05, -0.25) {\small$\uparrow$};
\node at (3.05, -0.65) {$\texttt{j}$};
\node at (4.05, -0.25) {\small$\uparrow$};
\node at (4.05, -0.65) {$\texttt{k}$};
\node at (4.85, -0.25) {\small$\uparrow$};
\node at (4.85, -0.65) {$\texttt{k} + \texttt{c}$};
\draw [decorate, decoration={brace,amplitude=5pt},yshift=2pt](4,0.5)--(5.5,0.5) node [midway,yshift=15pt]{$B$};
\draw (6.25,-1) rectangle (8.25, -0.5);
\draw (6.25, -1) -- (6.25, -0.5);
\draw (6.75, -1) -- (6.75, -0.5);
\draw (7.25, -1) -- (7.25, -0.5);
\draw (7.75, -1) -- (7.75, -0.5);
\draw (8.25, -1) -- (8.25, -0.5);
\node at (6.5, -0.75) {1};
\node at (7, -0.75) {2};
\node at (7.5, -0.75) {3};
\node at (7.5, -1.25) {\small$\uparrow$};
\node at (7.5, -1.6) {$\texttt{num}_{\leq \texttt{q}}$};
\node at (6.25, -1.3) {$\texttt{block}$};

\begin{scope}[yshift = -3cm]

\draw[fill=gruen!30] (-0.5, 0) rectangle (0, 0.5);
\draw[fill=hellgruen!30] (0,0) rectangle (1, 0.5);
\draw[fill=gruen!30] (1,0) rectangle (3, 0.5);
\draw[fill=dunkelgruen!30] (3.75, 0) rectangle (5.5, 0.5);
\draw[fill=gruen!30] (7.5,0) rectangle (8, 0.5);
\fill[fill=hellgruen!30] (5.5,0) rectangle (7.5, 0.16);
\fill[fill=gruen!30] (5.5,0.16) rectangle (7.5, 0.33);
\fill[fill=dunkelgruen!30] (5.5,0.33) rectangle (7.5, 0.5);
\fill[fill=hellgruen!30] (3, 0) rectangle (3.25, 0.5);
\fill[fill=hellgruen!30] (3.25,0.0) rectangle (3.75, 0.25);
\fill[fill=gruen!30] (3.25,0.25) rectangle (3.75, 0.5);

\draw (0,0) rectangle (8, 0.5);
\draw (7.5, 0) -- (7.5, 0.5);
\node at (-0.25, 0.25) {$p$};
\node at (7.75, 0.25) {$q$};
\node at (0.5, 0.25) {$< p$};
\node at (2, 0.25) {$p \leq \dots \leq q$};
\node at (4.65, 0.25) {$> q$};
\node at (6.5, 0.25) {? $\cdots$ ?};
\node at (1.05, -0.25) {\small$\uparrow$};
\node at (1.05, -0.65) {$\texttt{i}$};
\node at (3.05, -0.25) {\small$\uparrow$};
\node at (3.05, -0.65) {$\texttt{j}$};
\node at (5.55, -0.25) {\small$\uparrow$};
\node at (5.55, -0.65) {$\texttt{k}$};
\draw[->] (3.6, -0.75) -- (3.35, - 0.1);
\node at (3.6, -1) {$\texttt{j} + \texttt{c}$};
\draw [decorate, decoration={brace,amplitude=5pt},yshift=2pt](3,0.5)--(3.75,0.5) node [midway,yshift=15pt]{$\texttt{num}_{\leq \texttt{q}}$};
\draw (6.25,-1) rectangle (8.25, -0.5);
\draw (6.25, -1) -- (6.25, -0.5);
\draw (6.75, -1) -- (6.75, -0.5);
\draw (7.25, -1) -- (7.25, -0.5);
\draw (7.75, -1) -- (7.75, -0.5);
\draw (8.25, -1) -- (8.25, -0.5);
\node at (6.5, -0.75) {0};
\node at (7, -0.75) {1};
\node at (7.0, -1.25) {\small$\uparrow$};
\node at (7.0, -1.7) {$\texttt{num}_{< \texttt{p}}$};
\node at (6.25, -1.3) {$\texttt{block}$};
\end{scope}
\end{tikzpicture}}
\vspace*{-2em}
\caption{Lomuto block partitioning with two pivots. Top: Algorithm~\ref{algo:dual:pivot:partitioning}
compares elements with $q$ and is on Line~8 with $\texttt{c} = 3$, cf.~Figure~\ref{fig:lomuto:block:invariant}. Bottom: Algorithm compares the $\texttt{num}_{\leq \texttt{q}}$ elements at most as large as $q$ to $p$. Algorithm is on Line~14 with $\texttt{c} = 1$.}
\label{fig:lomuto:two:block:invariant}
\end{figure}

\subsection{Two Pivots}

Algorithm~\ref{algo:dual:pivot:partitioning} (\lomutotwo) describes a two-pivot version of Algorithm~\ref{algo:single:pivot:partitioning}, see Figure~\ref{fig:lomuto:two:block:invariant}. Compared to the latter, it uses one more index to store the beginning of a segment and one more \texttt{num} variable to store the number of misplaced elements in a block. The difference to \lomutoone~lies in the Lines~12--18 of Algorithm~\ref{algo:dual:pivot:partitioning}. Directly after moving misplaced elements (with respect to $q$) into a consecutive segment of the array, the algorithm checks the $\texttt{num}_{\leq \texttt{q}}$ elements in this segment immediately to $p$ using the same block to store misplaced elements. This can be done since all misplaced elements w.r.t.\ $q$ have been moved. It then moves misplaced elements smaller than $p$ in the very same fashion. After the rearrangement phase ends, the two pivots are swapped into place and their position is returned.

\subsection{Handling Equal Elements}
\label{sec:equal:elements}
 Algorithm~\ref{algo:dual:pivot:partitioning} ensures that all elements equal to $p$ or $q$ are stored between these pivots in the resulting partition. Thus, if $p$ equals $q$, the call on Line~5 of Algorithm~\ref{algo:dual:pivot:qs} can be avoided. So, from now on, Line~5 is guarded by the statement \texttt{if (p != q)}.

\section{Theoretical Properties of the Algorithms}\label{sec:theory}

In this section we analyze the theoretical properties of the proposed algorithms with respect to the number of element comparisons they make and their memory access pattern. The analysis is provided in a general way that allows us to choose the pivot(s) from a sample of the array elements.

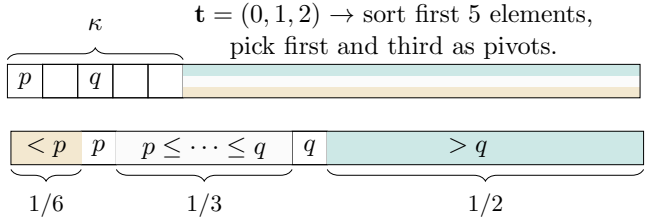
\begin{figure}[t!]
\centering
\scalebox{.89}{
\begin{tikzpicture}[xscale = 1.05, xshift=0.5cm]
\node[align=center] at (5, 1) {$\mathbf{t} = (0, 1, 2) \rightarrow$ sort first 5 elements,\\ pick first and third as pivots.};
\draw (-0.5, 0) rectangle (0, 0.5);
\draw (-0, 0) rectangle (0.5, 0.5);
\draw (0.5, 0) rectangle (1, 0.5);
\draw (1.0, 0) rectangle (1.5, 0.5);
\draw(1.5, 0) rectangle (2.0, 0.5);
\fill[fill=hellgruen!30] (2, 0) rectangle (8.5, 0.16);
\fill[fill=gruen!30] (2, 0.16) rectangle (8.5, 0.33);
\fill[fill=dunkelgruen!30] (2, 0.33) rectangle (8.5, 0.5);

\draw (0,0) rectangle (8.5, 0.5);

\node at (-0.25, 0.25) {$p$};
\node at (0.75, 0.25) {$q$};

\draw [decorate, decoration={brace,amplitude=5pt},yshift=2pt](-0.5,0.5)--(2,0.5) node [midway,yshift=15pt]{$\kappa$};

\begin{scope}[yshift = -1cm, xshift = -0.45cm]

\draw (1, 0) rectangle (1.5, 0.5);
\draw (4.0, 0) rectangle (4.5, 0.5);
\fill [fill=hellgruen!30] (0, 0) rectangle (1, 0.5);
\fill [fill=gruen!30] (1.5, 0) rectangle (4, 0.5);
\fill [fill=dunkelgruen!30] (4.5, 0) rectangle (9, 0.5);
\node at (1.25, 0.25) {$p$};
\node at (0.5, 0.25) {$ < p$};
\node at (2.75, 0.25) {$ p \leq \dots \leq q$};
\node at (4.25, 0.25) {$q$};
\node at (6.5, 0.25) {$> q$};
\draw [decorate, decoration={brace,amplitude=5pt},yshift=-2pt](1,0.0)--(0,0.0) node [midway,yshift=-15pt]{\small $1/6$};
\draw [decorate, decoration={brace,amplitude=5pt},yshift=-2pt](4,0.0)--(1.5,0.0) node [midway,yshift=-15pt]{\small $1/3$};
\draw [decorate, decoration={brace,amplitude=5pt},yshift=-2pt](9,0.0)--(4.5,0.0) node [midway,yshift=-15pt]{\small $1/2$};
\draw (0,0) rectangle (9, 0.5);
\end{scope}
\end{tikzpicture}}
\vspace*{-2.5em}
\caption{Top: Sampling step with $\mathbf{t} = (0,1,2)$ and two pivots. 
Bottom: Assuming a random permutation, expected sizes of different groups under the pivot choice. The 5 samples split the input into 6 different parts of equal size (in expectation). Thus, a fraction of $1/6$ of the remaining elements are smaller than $p$, $1/3$ are in between $p$ and $q$, and $1/2$ are larger than $q$.}
\label{fig:sampling:step}
\vspace{-1em}
\end{figure}

\subsection{Setup of the Analysis}

The input is a random permutation
of the set $\{1,\ldots,n\}$ which resides in an array $A[1..n]$.  Fix an
integer $k \in \{1, 2\}$ which denotes the number of pivots. 
Fix a vector $\mathbf{t} = (t_0, \ldots,
t_k) \in \mathbb{N}^{k+1}$. Let $\kappa := \kappa(\mathbf{t}) = k + \sum_{0 \leq i \leq k} t_i$ be
the number of samples. We assume that $\kappa$ is a constant independent of $n$. The general outline of a $k$-pivot quicksort algorithm is then as follows: If $n \leq \kappa$, sort $A$ directly. Otherwise, sort 
the first $\kappa$ elements and then set $p_i = A[i + \sum_{j < i} t_j]$, for $1
\leq i \leq k$. Next, partition the input $A[\kappa + 1..n]$ with respect to the
pivots $p_1,\ldots,p_k$. Subsequently, by a constant number of
swaps, move the elements residing in $A[1..\kappa]$ to correct final locations. Finally, sort the $k + 1$ subproblems recursively. (Algorithm~\ref{algo:single:pivot:qs} and Algorithm~\ref{algo:dual:pivot:qs} have to be slightly adapted for this to hold.) See Figure~\ref{fig:sampling:step} for an example.

\subsection{Cost Measures}
We measure cost in
two ways: we count the number of \emph{comparisons} with the pivot(s)
and we count the number of array cells \emph{accessed} by the respective algorithm when sorting an input containing $n$ elements.

With regard to comparisons, we define the following random variables: 
Let $\text{PCMP}_n$ denote the number of
comparisons with elements against pivot $p$ in Algorithm~\ref{algo:single:pivot:partitioning}, i.e., the number of times Line~7 is reached. Let $\text{CMP}_n$ be the number of 
comparisons over the whole recursion. Let $\text{PCMP}'_n$ denote the number of comparisons
with elements against the pivots $p$ and $q$ in Algorithm~\ref{algo:dual:pivot:partitioning} (Line~7 and Line~12) and define $\text{CMP}'_n$ accordingly.

With regard to array accesses, we define the cost as the number of times a cell of the array has been accessed. (This is identical to the notion of \emph{scanned elements} used in \cite{Wild18}.) More precisely, for Algorithm~\ref{algo:single:pivot:partitioning}, the random variable $\text{PMA}_n$ counts the number of times we reach Line~7 (reading $A[\texttt{j} + \texttt{c}]$) and 
Line~9 (reading $A[\texttt{i}]$). For Algorithm~\ref{algo:dual:pivot:partitioning},  $\text{PMA}'_n$ counts the sum of the number of times Line~7, Line~12, and Line~14 are reached, i.e., individual accesses of $\texttt{i}$, $\texttt{j}$, and $\texttt{k}$ in the array. We let by $\text{MA}_n$ and $\text{MA}'_n$ denote the cost over the whole recursion, respectively. 

\paragraph{From Partitioning Cost to Sorting Cost} Computing the average sorting cost from a given average partitioning cost is straight-forward if the latter is bounded by $a \cdot n + O(1)$, see \cite{AumullerD16}. 

For one-pivot quicksort with pivot $p$, 
we let by $a_0 := p - 1, a_1 := n - p$ denote the number of elements smaller/larger than the pivot. For two-pivot quicksort with pivots $p$ and $q$, we set 
$a_0 := p - 1, a_1 := q - p, a_2 := n - q$ to denote group sizes in the partition.
For a given sequence $\mathbf{t} = (t_0, \ldots, t_k) \in \mathbb{N}^{k + 1}$ we define
$H(\mathbf{t})$ by
\begin{align}
    H(\mathbf{t}) = \sum_{i = 0}^{k} \frac{t_i + 1}{\kappa + 1} \left(H_{\kappa + 1} -
    H_{t_i + 1}\right),
    \label{eq:entropy}
\end{align}
where $H_{\ell}$ denotes the $\ell$-th harmonic number.
Let $\text{P}_n$ denote the random variable which counts the cost of a single partitioning step,
and let $\text{C}_n$ denote the cost over the whole sorting procedure. Following \cite{AumullerDK16}, the average sorting cost $\E(C_n)$ follows the recurrence
    \begin{align}
        \label{eq:sampling:recurrence}
        \E(\text{C}_n) &= \E(\text{P}_n) +\\ \sum_{a_0 + \cdots + a_k = n - k}& \!\!\!\!\!\!\!\!\!\!\!\left(
                       \E(\text{C}_{a_0}) + \cdots + \E(\text{C}_{a_k})\right) \cdot \Pr(\langle a_0, \ldots, a_k\rangle),\notag
        \end{align}
    where $\langle a_0,\ldots,a_k\rangle$ is the event that the group sizes are exactly $a_0, \ldots, a_k$.
        The probability of this event for a given vector $\mathbf{t}$ is
            $\frac{\binom{a_0}{t_0} \cdots \binom{a_k}{t_k}}{\binom{n}{\kappa}}$.
        Now, a result by Hennequin~\cite[Proposition III.9]{hennequin}  says that
        for fixed $k$ and $\mathbf{t}$ and average partitioning cost $\E(P_n) = a \cdot n + O(1)$ recurrence \eqref{eq:sampling:recurrence}
has the solution
\begin{align}
    \E(C_n) = \frac{a}{H(\mathbf{t})} n \ln n + O(n).
    \label{eq:quicksort:recurrence:sampling}
\end{align}

The sampling technique used here does not preserve randomness in subproblems, since a few elements have already been sorted during the pivot sampling step.
For the analysis, we ignore that the unused samples have been seen and get
only an estimate on the sorting cost. See~\cite{NebelWM15} for a detailed analysis of this situation.

\subsection{Analysis}

\paragraph*{\lomutoone~(Algorithm~\ref{algo:single:pivot:partitioning})} Every element in the input that has not been sampled is compared with the pivot exactly once (Line~7), so we get $\text{PCMP}_n = n + O(1)$. Conditioned on the pivot being $p$, there are exactly $p - 1$ elements that are smaller than $p$ in the input. Each of these elements is accessed exactly once in Line~9, thus we get $\text{PMA}_n = n + p  + O(1)$. If the pivot is chosen according to a vector $\mathbf{t} = (t_0, t_1)$, we expect that there are $\frac{t_0 + 1}{t_0 + t_1 + 2} \cdot n + O(1)$ many elements smaller than $p$. Thus, we get $\E(\text{PMA}_n) = \left(1 + \frac{t_0 + 1}{t_0 + t_1 + 2}\right) \cdot n + O(1)$ memory accesses on average.

\paragraph*{\lomutotwo~(Algorithm~\ref{algo:dual:pivot:partitioning})} Each unsampled element from the input is compared exactly once to $q$ (Line~7). Each element that is smaller than $q$ is then compared to $p$ (Line~12). If the pivots $p$ and $q$ are chosen according to $\mathbf{t} = (t_0, t_1, t_2) \in \mathbb{N}^3$, we expect that a fraction of $\frac{t_0 + t_1 + 2}{t_0 + t_1 + t_2 + 3}$ elements are smaller than $q$. Thus, we obtain $\E(\text{PCMP}'_n) = \left(1 + \frac{t_0 + t_1 + 2}{t_0 + t_1 + t_2 + 3}\right) \cdot n + O(1)$. Regarding memory accesses, fix the pivots $p, q$. Every array cell is accessed by Line~7 of the algorithm. All array cells $A[j]$ with $j < q$ are accessed a second time in Line~12. Finally, all array cells $A[j]$ with $j < p$ are accessed a third time in Line~14 of the algorithm. These considerations yield $\E(\text{PMA}'_n) = \left(1 + \frac{2t_0 + t_1 + 3}{t_0 + t_1 + t_2 + 3}\right) \cdot n + O(1).$ 

\begin{table}[h!]
\small{
\begin{tabular}{c c l l}
\hline
\textbf{Algo} & \textbf{AS} & \textbf{Cost} & \textbf{Best} (cost, \textbf{t}) \\
\hline
$H_1$ &  & cmp & $2.00 n \ln n$, $(0, 0)$\\
$H_1$ &  0 & ma & $2.00 n \ln n$, $(0, 0)$\\
$H_1$ &   & cmp + ma & $4.00 n \ln n$, $(0, 0)$\\[.5em]
$H_1$ &  & cmp & $1.71 n \ln n$, $(1, 1)$\\
$H_1$ & 2 & ma & $1.71 n \ln n$, $(1, 1)$\\
$H_1$ &   & cmp + ma & $3.43 n \ln n$, $(1, 1)$ \\[.5em]
$H_1$ &  & cmp & $1.62 n \ln n$, $(2, 2)$\\
$H_1$ & 4 & ma & $1.62 n \ln n$, $(2, 2)$\\
$H_1$ &   & cmp + ma & $3.24 n \ln n$, $(2, 2)$\\[.5em]
$H_1$ &  & cmp & $1.53 n \ln n$, $(5, 5)$ \\
$H_1$ & 10 & ma & $1.53 n \ln n$, $(5, 5)$\\
$H_1$ &   & cmp + ma & $3.06 n \ln n$, $(5, 5)$ \\[.5em]
\hline
$L_1$  &  & cmp & $2.00 n \ln n$, $(0, 0)$ \\
$L_1$  & 0 & ma & $3.00 n \ln n$, $(0, 0)$\\
$L_1$  &  & cmp + ma & $5.00 n \ln n$, $(0, 0)$ \\[.5em]
$L_1$  &  & cmp & $1.71 n \ln n$, $(1, 1)$\\
$L_1$  & 2 & ma & $2.57 n \ln n$, $(1, 1)$\\
$L_1$  &  & cmp + ma & $4.29 n \ln n$, $(1, 1)$\\[.5em]
$L_1$  &  & cmp & $1.62 n \ln n$, $(2, 2)$\\
$L_1$  & 4 & ma & $2.38 n \ln n$, $(1, 3)$\\
$L_1$  &  & cmp + ma & $4.05 n \ln n$, $(2, 2)$ \\[.5em]
$L_1$  &  & cmp & $1.53 n \ln n $, $(5, 5)$ \\
$L_1$  & 10 & ma & $2.22 n \ln n$, $(4, 6)$\\
$L_1$  &   & cmp + ma & $3.78 n \ln n$, $(4, 6)$ \\[.5em]

\hline
$L_2$ &  & cmp & $2.00 n \ln n$, $(0, 0, 0)$ \\
$L_2$ & 0 & ma &  $2.40 n \ln n$, $(0, 0, 0)$ \\
$L_2$ &  & cmp + ma & $4.40 n \ln n$, $(0, 0, 0)$ \\[.5em]
$L_2$ &  & cmp & $1.73 n \ln n$, $(0, 1, 2)$\\
$L_2$ & 3 & ma & $1.92 n \ln n$, $(0, 1, 2)$ \\
$L_2$ &  & cmp + ma & $3.65 n \ln n$, $(0, 1, 2)$  \\[.5em]
$L_2$ &  & cmp & $1.62 n \ln n$, $(1, 1, 3)$\\
$L_2$ & 5 & ma & $1.88 n \ln n$, $(0, 2, 3)$ \\
$L_2$ &  & cmp + ma & $3.51 n \ln n$, $(1, 1, 3)$  \\[.5em]
$L_2$ &  & cmp & $1.55 n \ln n$, $(2, 3, 6)$ \\
$L_2$ & 11 & ma & $1.77 n \ln n$, $(1, 3, 7)$ \\
$L_2$ &   & cmp + ma & $3.32 n \ln n$, $(2, 3, 6)$ \\[.5em]
\end{tabular}}
\caption{Best asymptotic expected sorting cost of Hoare's one-pivot ($H_1$), Lomuto's one-pivot ($L_1$), and Lomuto's two-pivot ($L_2$) algorithm with regard to a given sample size AS (in addition to the pivot(s)). For each cost measure (cmp--comparisons, ma--memory accesses), we explicitly state the sample vector $\textbf{t}$ for which this cost is achieved.}
\label{tab:sorting:cost}
\end{table}
\paragraph{Comparison with One-Pivot Hoare Partitioning}

BlockQuicksort~\cite{EdelkampW16} uses Hoare's partitioning scheme. It is immediate that 
in Hoare's partitioning scheme every element is compared once to the pivot, and every array cell is accessed exactly once, as well.
\vspace{-.6em}
\subsection{Comparison to \ips} 

\ips~with $2^\ell$ pivots makes $\ell$ comparisons per element. Thus, it makes $\ell \cdot n + O(1)$ comparisons in the partitioning step. A rough estimate for the number of array accesses is obtained by counting three array accesses per array position; the first during classification, a second when a block is full and is moved back into the array, the third when a block is moved to a final position during the final rearrangement phase.
\vspace{-.6em}

\subsection{Putting Everything Together} Using \eqref{eq:quicksort:recurrence:sampling} with the cost formulas derived above, we can reason about the expected sorting cost of Algorithms~\ref{algo:single:pivot:partitioning}/\ref{algo:dual:pivot:partitioning}, and compare them to the two other approaches. Table~\ref{tab:sorting:cost} gives an overview over the minimum cost achievable with certain sample sizes; we present a short summary.

First, Lomuto's partitioning scheme is inferior to Hoare's, in particular
with regard to the number of memory accesses. However, choosing pivots from a sample greatly improves the number of accesses, both for the one- and the two-pivot variant. While \lomutotwo~makes slightly more comparisons, the number of memory accesses improves by around $0.5 n \ln n$ for the sample sizes considered. The pivot choices that achieve the minimum cost are the median of the sample (Hoare's scheme), the median or the element one larger than it (\lomutoone), and a skewed pivot choice for \lomutotwo. E.g., $(1,1,1)$ accesses around $9\%$ more array cells than $(0,1,2)$. As a rule of thumb, the calculations motivate to take the larger pivot $q$ as the median in a sample, and to choose the smaller pivot $p$ as the median of all the sampled elements smaller than $q$. For a sample size of 5, \lomutotwo~makes a factor of $1.18$ more memory accesses than Hoare.

Using 128 pivots from a sample of 257 elements, taking every second element as a pivot, \ips~outperforms its competitors by a large margin. (The same behavior was observed for other samplesort-based approaches in \cite{Aumuller15}.) On average, it makes around $1.51 n \ln n + O(n)$ comparisons and $0.65n \ln n + O(n)$ memory accesses. In comparison, \lomutotwo~with a sample of 257 elements makes $1.45 n\ln n + O(n)$ comparisons, but incurs $1.69 n \ln n + O(n)$ memory accesses, on average.  


\section{Experimental Evaluation}\label{sec:evaluation}

This section presents the evaluation of the experiments we conducted with regard to the 
performance and other properties of the proposed algorithms. The implementation was written in C++, performance counters were obtained using PAPI\footnote{\url{http://icl.utk.edu/papi/}}.
Our code, the Jupyter notebook that makes all computations transparent and which includes additional plots not found in the paper, as well as raw results are available at \url{https://bitbucket.org/alenex19_paper48/submission/}.

\paragraph*{Input Distributions}

We ran benchmarks with the following input distributions: \textsf{Permutation}, \textsf{Sawtooth}, \textsf{RandomDup}, \textsf{Sorted}, \textsf{Reversed}, \textsf{Equal}, and \textsf{EightDup}. \textsf{Permutation} chooses the input as a random permutation of $\{1, \ldots, n\}$; \textsf{Sawtooth} sets $A[i] = i \mod \sqrt{n}$, \textsf{RandomDup} uses $A[i] = \texttt{uniform($n$)} \mod \sqrt{n}$; \textsf{Sorted} sets $A[i] = i$, \textsf{Reversed} sets $A[i] = n - i - 1$, and \textsf{Equal} uses
$A[i] = 1$. Finally, \textsf{EightDup} sets $A[i] = i^8 + n / 2 \mod n$, following \cite{EdelkampW16}, resulting in inputs that duplicate the median many times when $n$ is a power of two. All inputs consisted of $64$-bit integers. For each input size, we ran each algorithm on the same $600$ inputs drawn from a certain input distribution.
All figures in the following contain the average over these trials. 
We call a running 
time improvement \emph{significant} if it was observed in at least 
95\% of the trials.

\paragraph*{Machine Details} We ran the experiments on two different machines: $\texttt{Xeon}$ and $\texttt{i7}.$ \texttt{Xeon} is set up with two 14-core Intel Xeon E5-2690 v4 CPUs clocked at 2.6 GHz, 35MB L3 Cache and 512GB of RAM. \texttt{Xeon} was running Ubuntu 16.10 with Linux kernel 4.4 and the code was compiled with \texttt{gcc} version 5.4.0. The compiler flags were \texttt{-O3 -march=native -funroll-loops}. All runs used a single core and a single thread.  There are marginal differences to results obtained on \texttt{i7}, thus we moved the discussion for this architecture to Appendix~\ref{sec:running:times:i7}. 

\paragraph*{Competitors.} To compare the running time to other implementations, we used the implementation of BlockQuicksort~\cite{EdelkampW16} (\blockqs) available at \url{https://github.com/weissan/BlockQuicksort} and the implementation of \ips~\cite{AxtmannWF017} retrieved from \url{https://github.com/SaschaWitt/ips4o}. Furthermore, we used \textsf{std::sort} from \texttt{gcc} as the baseline implementation to compare all algorithms to. The implementations of \lomutoone~and \lomutotwo~save lines of code by a factor of 5 and 8, resp., compared to \blockqs~under a similarly dense coding style.

\subsection{Block Size}

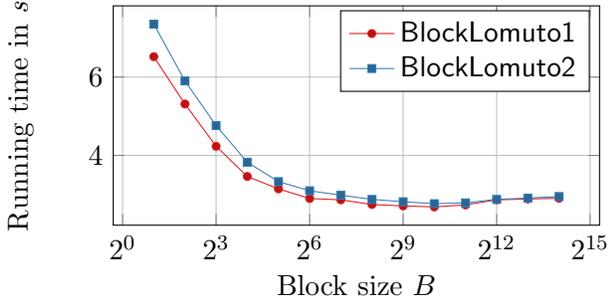
\begin{figure}[t!]
\begin{tikzpicture}[every mark/.append style={mark size=1.5pt}]
                \begin{axis}[
                    height = 4.5cm,
                    width = .95\columnwidth,
                    xlabel = {Block size $B$},
                    ylabel = {Running time in \emph{s}},
                    xmode=log,
                    log basis x = 2,
                    grid = both, grid style={line width=.1pt, draw=gray!30},
    major grid style={line width=.2pt,draw=gray!50},
                ]
\addplot coordinates { (2.000000, 6.522000) (4.000000, 5.313000) (8.000000, 4.233000) (16.000000, 3.462000) (32.000000, 3.145000) (64.000000, 2.900000) (128.000000, 2.865000) (256.000000, 2.746000) (512.000000, 2.710000) (1024.000000, 2.684000) (2048.000000, 2.736000) (4096.000000, 2.865000) (8192.000000, 2.887000) (16384.000000, 2.906000) };
\addlegendentry{\lomutoone};
\addplot coordinates { (2.000000, 7.353000) (4.000000, 5.903000) (8.000000, 4.762000) (16.000000, 3.825000) (32.000000, 3.331000) (64.000000, 3.097000) (128.000000, 2.983000) (256.000000, 2.877000) (512.000000, 2.817000) (1024.000000, 2.767000) (2048.000000, 2.787000) (4096.000000, 2.877000) (8192.000000, 2.910000) (16384.000000, 2.951000) };
\addlegendentry{\lomutotwo};
\end{axis}\end{tikzpicture}
\caption{Influence of block size $B$ to the running time for $n=2^{26}$, \texttt{Xeon} on \textsf{Permutation}.}
\label{fig:plot:blocksize}
\end{figure}

Figure~\ref{fig:plot:blocksize} shows how the block size influences the sorting time for 
the one- and two-pivot variant for $2 \leq B \leq 2^{14}$. Both variants benefit from large block sizes, decreasing the running time to sort the input of $n = 2^{26}$ items from around 7 seconds to around 3 seconds. The minimum is attained for both variants for a block that can hold up to $1024$ items, but there is no big difference for block sizes around this value. Consequently, we set the block size to $B = 1024$ for both implementations. 

\subsection{Pivot Strategies}

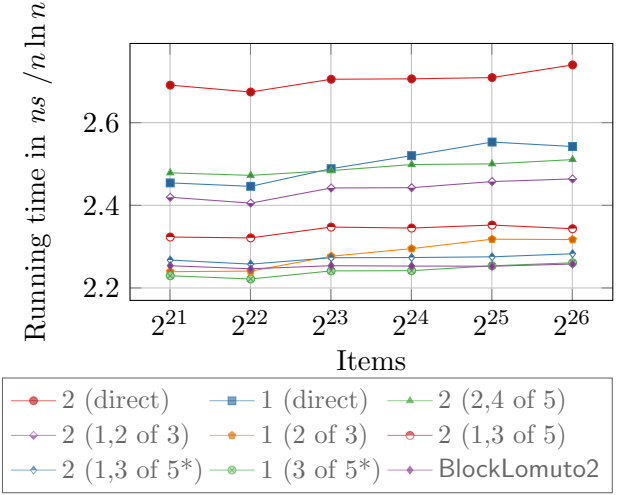
\begin{figure}[t!]
\begin{tikzpicture}[every mark/.append style={mark size=1.5pt}]
                \begin{axis}[
                    xlabel = {Items},
                    ylabel = {Running time in \emph{ns} $/ n \ln n$},
                    xmode=log,
                    legend style = {opacity = 0.6, text opacity = 1, at = { (1, -0.3)}, font = \small},
                    legend columns = 3,
                    legend cell align = {left},
                    log basis x = 2,
                    width = 8cm, 
                    height = 5cm,
                    grid = both, grid style={line width=.1pt, draw=gray!30},
    major grid style={line width=.2pt,draw=gray!50},
                ]
\addplot coordinates { (2097152.000000, 2.690791) (4194304.000000, 2.674174) (8388608.000000, 2.705063) (16777216.000000, 2.706075) (33554432.000000, 2.709071) (67108864.000000, 2.739816) };
\addlegendentry{2 (direct)};    
\addplot coordinates { (2097152.000000, 2.454274) (4194304.000000, 2.445906) (8388608.000000, 2.488664) (16777216.000000, 2.520191) (33554432.000000, 2.553082) (67108864.000000, 2.542284) };
\addlegendentry{1 (direct)};   
\addplot coordinates { (2097152.000000, 2.478516) (4194304.000000, 2.472485) (8388608.000000, 2.484028) (16777216.000000, 2.498693) (33554432.000000, 2.500284) (67108864.000000, 2.510616) };
\addlegendentry{2 (2,4 of 5)};   
\addplot coordinates { (2097152.000000, 2.419550) (4194304.000000, 2.405256) (8388608.000000, 2.442004) (16777216.000000, 2.442655) (33554432.000000, 2.457460) (67108864.000000, 2.463900) };
\addlegendentry{2 (1,2 of 3)};
\addplot coordinates { (2097152.000000, 2.239378) (4194304.000000, 2.240778) (8388608.000000, 2.276751) (16777216.000000, 2.295180) (33554432.000000, 2.317982) (67108864.000000, 2.317135) };
\addlegendentry{1 (2 of 3)};
\addplot coordinates { (2097152.000000, 2.323240) (4194304.000000, 2.321140) (8388608.000000, 2.347339) (16777216.000000, 2.344983) (33554432.000000, 2.352207) (67108864.000000, 2.343346) };
\addlegendentry{2 (1,3 of 5)};
\addplot coordinates { (2097152.000000, 2.267550) (4194304.000000, 2.257663) (8388608.000000, 2.273162) (16777216.000000, 2.273611) (33554432.000000, 2.275330) (67108864.000000, 2.282739) };
\addlegendentry{2 (1,3 of 5*)};
\addplot coordinates { (2097152.000000, 2.229550) (4194304.000000, 2.221703) (8388608.000000, 2.241308) (16777216.000000, 2.241794) (33554432.000000, 2.253661) (67108864.000000, 2.260910) };
\addlegendentry{1 (3 of 5*)};
\addplot coordinates { (2097152.000000, 2.253792) (4194304.000000, 2.246406) (8388608.000000, 2.253870) (16777216.000000, 2.252973) (33554432.000000, 2.252629) (67108864.000000, 2.257851) };
\addlegendentry{\lomutotwo};
\end{axis}\end{tikzpicture}
\caption{Running times on \textsf{Permutation} on \texttt{Xeon} for different pivot choices. Legend entries of the form $\texttt{p (R of S)}$ are read as follows: For the \texttt{p}-pivot variant, we sort \texttt{S} elements and choose the elements \texttt{R} in the sorted sample, e.g., 
\texttt{(2, 4 of 5)} means that the second and the fourth element in a sorted sample of five elements are chosen as pivots; a \texttt{*} refers to the median of medians sampling.}
\label{fig:plot:pivot:choice}
\end{figure}

We implemented different pivot selection strategies and compared them to each other. Figure~\ref{fig:plot:pivot:choice} shows the result of this experiment. For \lomutoone, we compare direct choice, median of 3, and median of medians of $5$, which groups $25$ elements in 5 groups, chooses the median in each, and chooses the median of these medians as pivot, see~\cite{EdelkampW16}. For \lomutotwo, we compare direct choice, first two of three elements, first and third of 5, second and fourth of 5, and the adaption of the median of medians strategy described above, in which we take the first and third element in the sample of 5 medians.

The plot shows that the pivot selection strategy has big influence on the running time. In particular, choosing the pivot directly from the input without sampling performs 
much worse than even simple pivot selection strategies. For example, 
for \lomutoone, there is a difference of a factor $1.14$ in running time when choosing the pivot directly from the input compared to choosing it as the median of 5 medians.
\lomutotwo~is a factor of $1.16$ times slower without sampling than choosing the first two elements in a sample of three elements, i.e., looking at one additional element. It is by a factor of $1.22$ slower when pivots are chosen among the medians of five elements each. In particular, 
choosing skewed pivots as predicted in Section~\ref{sec:theory} provides an improvement for 
the two-pivot variant. Choosing the second and fourth element is a factor of $1.07$ slower than choosing the first and third element.

For the final implementation, we used a slightly more elaborate pivot selection strategy that switches between pivot strategies based on the input size. The best thresholds to switch strategies have been obtained experimentally as well. While improvements in running time are consistent, they do not exceed a factor of $1.01$ compared to the fastest variant that does not switch strategies, as examplified through the difference between $\lomutotwo$ and \textsf{2 (1, 3 of 5$^*$)} in
Figure~\ref{fig:plot:pivot:choice}.

\subsection{Further Tuning}

Apart from choosing the block size and pivot selection strategy based on the experiments described above, the implementations used in the experiments are identical to Algorithm~\ref{algo:single:pivot:partitioning} and Algorithm~\ref{algo:dual:pivot:partitioning}. In contrast to the observations made in \cite[Section~3.2]{EdelkampW16}, unrolling the main loop did not increase performance. We suspect that this is due to the simplicity of the code that allows the compiler to unroll the code. Furthermore, we do not perform the cyclic rotations described in \cite{EdelkampW16} since they actually increased running time. We suspect that this is due to the CPU pipelining the load/store instruction issued in Line~9 of Algorithm~\ref{algo:single:pivot:partitioning}, whereas the cyclic shifts described in \cite{EdelkampW16} introduce read/write dependencies.

\subsection{Running Times}

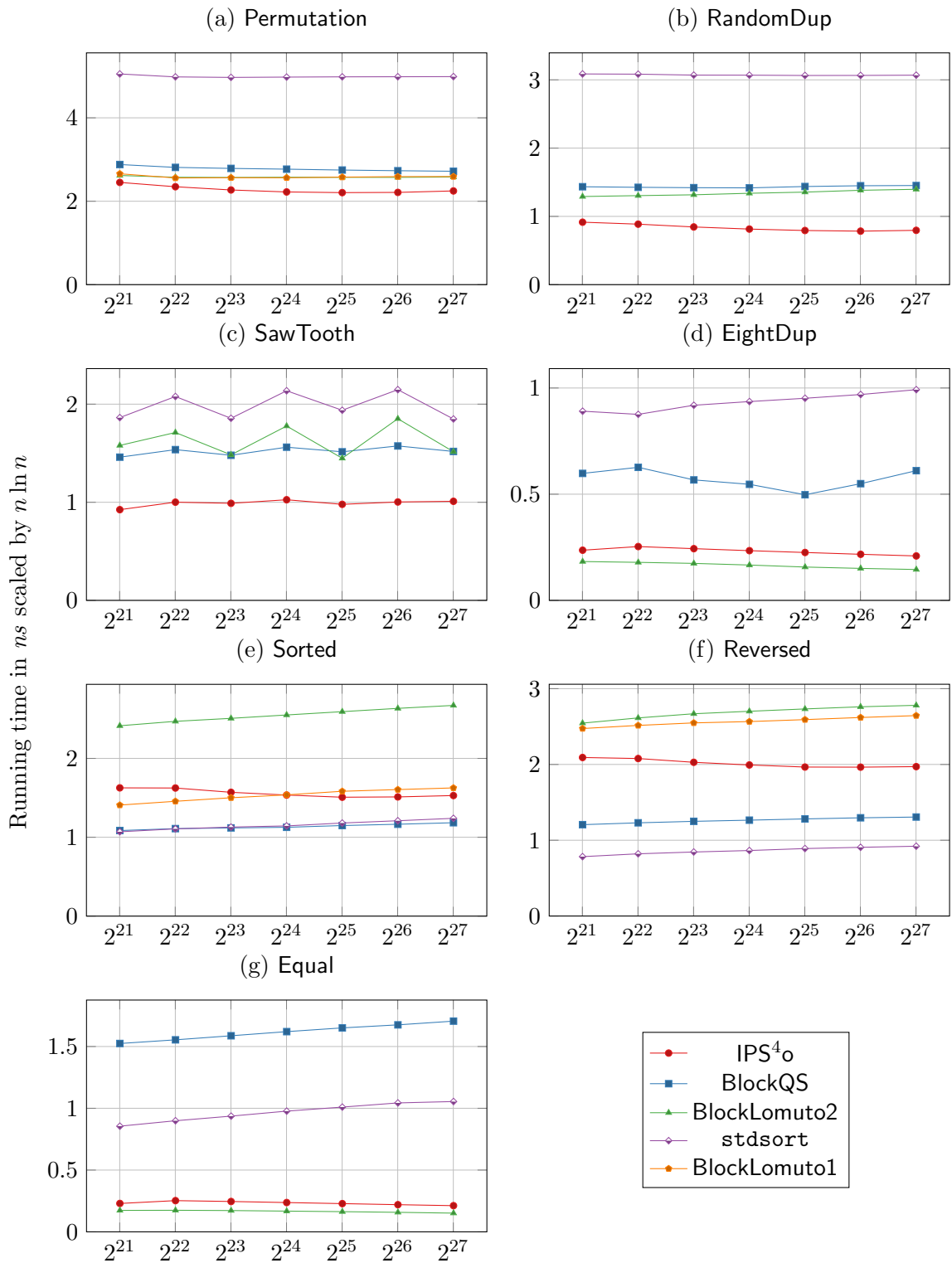
\begin{figure*}[t!]
\centering
\begin{tikzpicture}[every mark/.append style={mark size=1.5pt}]

\begin{groupplot}[group style = { group size = 2 by 4, group name = preciousss, vertical sep = 3.5 em},
                    height = 5.3cm,
                    width = 8cm,
                    xlabel = {},
                    ylabel = {},
                    ymin = 0,
                    xmode=log,
                    log basis x = 2,
                    grid = both, grid style={line width=.1pt, draw=gray!30},
    major grid style={line width=.2pt,draw=gray!50},
                ]

\nextgroupplot[
                    title = {(a) \textsf{Permutation}},
]

\addplot coordinates { (2097152.000000, 2.4512325) (4194304.000000, 2.3479429) (8388608.000000, 2.2703740) (16777216.000000, 2.2228604) (33554432.000000, 2.2067782) (67108864.000000, 2.2118753) (134217728.000000, 2.2471251) };
\addplot coordinates { (2097152.000000, 2.8800893) (4194304.000000, 2.8118459) (8388608.000000, 2.7870590) (16777216.000000, 2.7692841) (33554432.000000, 2.7478356) (67108864.000000, 2.7318499) (134217728.000000, 2.7188467) };
\addplot coordinates { (2097152.000000, 2.6177865) (4194304.000000, 2.5749531) (8388608.000000, 2.5696558) (16777216.000000, 2.5746571) (33554432.000000, 2.5733027) (67108864.000000, 2.5723537) (134217728.000000, 2.5824628) };
\addplot coordinates { (2097152.000000, 5.0522185) (4194304.000000, 4.9814577) (8388608.000000, 4.9691770) (16777216.000000, 4.9767172) (33554432.000000, 4.9828067) (67108864.000000, 4.9853400) (134217728.000000, 4.9880434) };
\addplot coordinates { (2097152.000000, 2.6588071) (4194304.000000, 2.5579932) (8388608.000000, 2.5638331) (16777216.000000, 2.5630952) (33554432.000000, 2.5782857) (67108864.000000, 2.5901354) (134217728.000000, 2.5920123) };
                \coordinate (top) at (rel axis cs:0,1);
\nextgroupplot[
                    title = {(b) \textsf{RandomDup}},
                ]
                
\addplot coordinates { (2097152.000000, .9149009) (4194304.000000, .8860680) (8388608.000000, .8450330) (16777216.000000, .8141792) (33554432.000000, .7932503) (67108864.000000, .7839763) (134217728.000000, .7955904) };
\addplot coordinates { (2097152.000000, 1.4340311) (4194304.000000, 1.4259372) (8388608.000000, 1.4203951) (16777216.000000, 1.4177564) (33554432.000000, 1.4373622) (67108864.000000, 1.4488345) (134217728.000000, 1.4521518) };
\addplot coordinates { (2097152.000000, 1.2908760) (4194304.000000, 1.3040529) (8388608.000000, 1.3163509) (16777216.000000, 1.3386034) (33554432.000000, 1.3554542) (67108864.000000, 1.3823519) (134217728.000000, 1.3968955) };
\addplot coordinates { (2097152.000000, 3.0864217) (4194304.000000, 3.0832469) (8388608.000000, 3.0694737) (16777216.000000, 3.0685544) (33554432.000000, 3.0638783) (67108864.000000, 3.0647781) (134217728.000000, 3.0684748) };
\nextgroupplot[
                    title = {(c) \textsf{SawTooth}},
                ]
                
\addplot coordinates { (2097152.000000, .9245881) (4194304.000000, 1.0008497) (8388608.000000, .9893489) (16777216.000000, 1.0256001) (33554432.000000, .9788638) (67108864.000000, 1.0029813) (134217728.000000, 1.0097730) };
\addplot coordinates { (2097152.000000, 1.4595360) (4194304.000000, 1.5362518) (8388608.000000, 1.4790294) (16777216.000000, 1.5612186) (33554432.000000, 1.5143809) (67108864.000000, 1.5739531) (134217728.000000, 1.5178754) };
\addplot coordinates { (2097152.000000, 1.5792921) (4194304.000000, 1.7110264) (8388608.000000, 1.4833343) (16777216.000000, 1.7769749) (33554432.000000, 1.4485534) (67108864.000000, 1.8514263) (134217728.000000, 1.5152661) };
\addplot coordinates { (2097152.000000, 1.8627008) (4194304.000000, 2.0778184) (8388608.000000, 1.8556714) (16777216.000000, 2.1362856) (33554432.000000, 1.9378637) (67108864.000000, 2.1475129) (134217728.000000, 1.8486292) };
\nextgroupplot[
                    title = {(d) \textsf{EightDup}}
                ]
                
\addplot coordinates { (2097152.000000, .2362363) (4194304.000000, .2534399) (8388608.000000, .2435210) (16777216.000000, .2341267) (33554432.000000, .2255209) (67108864.000000, .2170253) (134217728.000000, .2093240) };
\addplot coordinates { (2097152.000000, .5979380) (4194304.000000, .6264636) (8388608.000000, .5672864) (16777216.000000, .5464851) (33554432.000000, .4973935) (67108864.000000, .5494515) (134217728.000000, .6107864) };
\addplot coordinates { (2097152.000000, .1829334) (4194304.000000, .1791076) (8388608.000000, .1741832) (16777216.000000, .1664444) (33554432.000000, .1570103) (67108864.000000, .1506325) (134217728.000000, .1454010) };
\addplot coordinates { (2097152.000000, .8907063) (4194304.000000, .8758161) (8388608.000000, .9187505) (16777216.000000, .9360719) (33554432.000000, .9515947) (67108864.000000, .9693077) (134217728.000000, .9920527) };
\nextgroupplot[
                    title = {(e) \textsf{Sorted}},
                ]

\addplot coordinates { (2097152.000000, 1.6262773) (4194304.000000, 1.6243426) (8388608.000000, 1.5699666) (16777216.000000, 1.5351038) (33554432.000000, 1.5076638) (67108864.000000, 1.5115125) (134217728.000000, 1.5287984) };
\addplot coordinates { (2097152.000000, 1.0850584) (4194304.000000, 1.1092658) (8388608.000000, 1.1182718) (16777216.000000, 1.1263584) (33554432.000000, 1.1483650) (67108864.000000, 1.1649013) (134217728.000000, 1.1840624) };
\addplot coordinates { (2097152.000000, 2.4137941) (4194304.000000, 2.4698495) (8388608.000000, 2.5077634) (16777216.000000, 2.5506409) (33554432.000000, 2.5926483) (67108864.000000, 2.6343218) (134217728.000000, 2.6718028) };
\addplot coordinates { (2097152.000000, 1.0693343) (4194304.000000, 1.1062728) (8388608.000000, 1.1289646) (16777216.000000, 1.1439150) (33554432.000000, 1.1804029) (67108864.000000, 1.2093865) (134217728.000000, 1.2400000) };
\addplot coordinates { (2097152.000000, 1.4080242) (4194304.000000, 1.4558829) (8388608.000000, 1.5008035) (16777216.000000, 1.5375510) (33554432.000000, 1.5829752) (67108864.000000, 1.6045425) (134217728.000000, 1.6256740) };
\nextgroupplot[
                    title = {(f) \textsf{Reversed}},
                    legend style = { at = { (0.75, -0.5)}}
                ]

\addplot coordinates { (2097152.000000, 2.0916366) (4194304.000000, 2.0781088) (8388608.000000, 2.0281882) (16777216.000000, 1.9931407) (33554432.000000, 1.9656389) (67108864.000000, 1.9643254) (134217728.000000, 1.9711674) };
\addlegendentry{\ips};
\addplot coordinates { (2097152.000000, 1.2053293) (4194304.000000, 1.2296760) (8388608.000000, 1.2492777) (16777216.000000, 1.2647481) (33554432.000000, 1.2813027) (67108864.000000, 1.2959283) (134217728.000000, 1.3058545) };
\addlegendentry{\blockqs};
\addplot coordinates { (2097152.000000, 2.5452496) (4194304.000000, 2.6138683) (8388608.000000, 2.6688930) (16777216.000000, 2.7014994) (33554432.000000, 2.7323989) (67108864.000000, 2.7606724) (134217728.000000, 2.7800814) };
\addlegendentry{\lomutotwo};
\addplot coordinates { (2097152.000000, .7833517) (4194304.000000, .8208486) (8388608.000000, .8447660) (16777216.000000, .8643101) (33554432.000000, .8901036) (67108864.000000, .9064277) (134217728.000000, .9209409) };
\addlegendentry{stdsort};
\addplot coordinates { (2097152.000000, 2.4738699) (4194304.000000, 2.5150686) (8388608.000000, 2.5481983) (16777216.000000, 2.5657010) (33554432.000000, 2.5914502) (67108864.000000, 2.6194581) (134217728.000000, 2.6444649) };
\addlegendentry{\lomutoone};
\nextgroupplot[
                    title = {(g) \textsf{Equal}},
                ]
                
\addplot coordinates { (2097152.000000, .2296846) (4194304.000000, .2526581) (8388608.000000, .2456254) (16777216.000000, .2370955) (33554432.000000, .2284249) (67108864.000000, .2198283) (134217728.000000, .2118833) };
\addplot coordinates { (2097152.000000, 1.5243981) (4194304.000000, 1.5540978) (8388608.000000, 1.5870260) (16777216.000000, 1.6203990) (33554432.000000, 1.6506895) (67108864.000000, 1.6753036) (134217728.000000, 1.7053627) };
\addplot coordinates { (2097152.000000, .1735270) (4194304.000000, .1745289) (8388608.000000, .1723138) (16777216.000000, .1677496) (33554432.000000, .1633049) (67108864.000000, .1579984) (134217728.000000, .1514255) };
\addplot coordinates { (2097152.000000, .8551399) (4194304.000000, .8991342) (8388608.000000, .9365684) (16777216.000000, .9770099) (33554432.000000, 1.0096291) (67108864.000000, 1.0433783) (134217728.000000, 1.0546455) };
                \coordinate (bot) at (rel axis cs:0,0);

\end{groupplot}
    \path (top-|current bounding box.west)-- node[anchor=south,rotate=90] {Running time in \emph{ns} scaled by $n \ln n$} (bot-|current bounding box.west);
\end{tikzpicture}
\caption{Running time plots on \texttt{Xeon}. The $x$-axis represents the number of items, the $y$-axis shows the running time in nanoseconds scaled by $n \ln n$.}
\label{plot:running:times:xeon}
\end{figure*}

We discuss the running times observed on \texttt{Xeon} plotted in Figure~\ref{plot:running:times:xeon}. In the following, running time differences are always stated for the data points associated with $2^{27}$ items.

On \textsf{Permutation}, \ips~has the fastest running time. On average, \lomutoone~and \lomutotwo~are a factor of $1.15$ times slower, \blockqs~is $1.20$ times slower; \stdsort~is a factor of $2.22$ slower than \ips.   With regard to significant running time differences, these factors decrease in absolute value by about $0.02$. 

For inputs containing many duplicates (\textsf{SawTooth}, \textsf{RandomDup}, \textsf{EightDup}, \textsf{Equal}), \lomutoone~cannot compete with the other algorithms; a drawback of Lomuto's partitioning scheme that had already been identified in \cite{EdelkampW16}. Thus, we omit it in the plots. In the respective order of these input types, it was $221$, $280$, $18571$, $21010$ times slower than \ips. On the other hand, \lomutotwo~is robust on all of these input types, being fastest on \textsf{RandomDup} and \textsf{Equal}, and being close in performance to \blockqs~on the other two. Only on \textsf{Sorted} and \textsf{Reversed} it is around a factor $2$ to $3$ slower than the fastest variant. For both of these input types, \stdsort~performs very well. Furthermore, we note that average running times predict the difference between implementations very well: Even for random structured inputs (\textsf{RandomDup}), significant differences that occurred in at least 95\% of the trials made differences only about a factor of $0.02$ smaller than average running time differences. 

\begin{table*}[t!]
\centering
\begin{tabular}{c r r r r}
\hline
\textbf{Algorithm} & \textbf{L1/L2 CM} & \textbf{CB (\% MP)} & \textbf{INS} &  \textbf{WOF (\%, cycles)} \\ \hline
\ips & .147 / .085 & 1.080 (\phantom{0}6.1\%) & 12.870 & \phantom{0}2.921 (41.3\%, \phantom{0}6.834) \\
\lomutoone & .169 / .143 & 0.926 (11.7\%) & 16.456 &  \phantom{0}3.001 (38.7\%, \phantom{0}7.758) \\
\lomutotwo & .144 / .121 & 0.884 (10.5\%) & 17.467 & \phantom{0}2.579 (32.7\%, \phantom{0}7.878) \\
\blockqs & .125 / .102 & 0.877 (14.7\%) & 17.566 & \phantom{0}3.240 (38.7\%, \phantom{0}8.354) \\
\stdsort & .137 / .115 & 2.128 (29.0\%) & 11.769 & 11.509 (74.0\%, 15.550) \\
\end{tabular}
\caption{CPU counter measurements for $n = 2^{27}$ items on 600 trials on \textsf{Permutation}. We use the abbreviations ``CM'' for ``Cache Misses'', ``INS'' for ``Instructions'', ``CB'' for ``conditional branch instructions'', ``MP'' for ``mispredicted'', and ``WOF'' for ``cycles without instruction finished''. All values are normalized by $n \ln n$.}
\label{tab:cpu:counter}
\end{table*}

Despite their simplicity, the \textsf{BlockLomuto} variants show very good performance. In particular, the two-pivot variant is robust on different input types, achieving robustness with the small twist introduced in Section~\ref{sec:algorithm}. On the other hand, \blockqs~achieves robustness with elaborate additional checks of the input after the main partitioning step finishes. 

\subsection{Practical Observations}

Comparing our running time results to the considerations in Section~\ref{sec:theory}, it is quite surprising that \lomutoone~and \lomutotwo~can compete in performance with \blockqs. In Table~\ref{tab:cpu:counter} we present selected measurements we got from running experiments. 
Experiments are run on \textsf{Permutation} to report on the average-case behavior of implementations. 

From Section~\ref{sec:theory}, we expect that \ips~and \blockqs~use fewer instructions than the Lomuto-variants discussed in this paper. Furthermore, they are expected to have a better cache behavior. 

From Table~\ref{tab:cpu:counter} we see that with regard to L1 cache misses, \blockqs~incurs the fewest misses, followed by \stdsort, \lomutotwo, \ips, and \lomutoone. So, \ips~uses too many blocks for all of them to reside in L1 cache. Looking at L2 cache misses, \ips~incurs fewer misses than all other algorithms, as predicted. \lomutotwo~has a better cache behavior than \lomutoone, but behaves worse than \blockqs, which again follows nicely the memory accesses computed in Section~\ref{sec:theory}.

\lomutoone~and \lomutotwo~branch conditionally in about the same dimension as 
\ips~and \blockqs, and there is a big difference to the number of branches in \stdsort. These branches are easy to predict, as shown by the low misprediction rates. Here, \textsf{LomutoBlock} variants make fewer mispredictions than \blockqs, and there is a big gap to the misprediction rate of \ips; \stdsort~mispredicts more branches, since comparisons to the pivot have a prediction rate of around 50\%.

With regard to the instruction count, we see that \ips~makes by far the fewest instructions among ``branch-free'' variants. Somehow surprisingly, the Lomuto-variants incur fewer instructions than \blockqs. We suspect that this is because of their simpler structure that simplifies bookkeeping. Looking at cycles in which no instruction was finished (because of, e.g.,  memory stalls or branch mispredictions), Lomuto-based variants have slightly fewer of them compared to \blockqs; around 33--40\% of the total amount of cycles are of this type for branch-free
variants, while they make up $74\%$ of the cycles for \stdsort.

In conclusion, we see that the theoretical differences between \ips/\blockqs~and the Lomuto-based variants translate into practice, but their easier structure make them competitive to \blockqs. No variant can compete with \ips, both in theory and in practice.

\vspace{-.6em}
\section{Conclusion}\label{sec:conclusion}

This paper introduced simple variants of the BlockQuicksort algorithm by~\cite{EdelkampW16} using block-based versions of Lomuto's partitioning scheme. 
The implementation was shown to be competitive in running time to the implementation of~\cite{EdelkampW16}. A novel twist to the general two-pivot quicksort approach made the proposed two-pivot variant particularly robust with regard to different input distributions. The paper presented theoretical properties of the algorithms and verified them through experiments.

Due to their simple structure, the proposed algorithms are particularly suited to test further optimizations. For example, it would be nice to inspect how much can be gained from implementing different block operations such as filling the block or rearranging elements through vectorized SIMD operations. In another line of research, having a simple implementation could allow to formally verify the correctness of the implementation. Additionally, a theoretical analysis of handling equal elements as described in Section~\ref{sec:equal:elements} seems interesting~\cite{Wild18b}. 

We remark that a 3-pivot variant of Lomuto's scheme did not give any improvements with regard to observed running times; the same is true for a two-pivot variant of Hoare's scheme used in BlockQuicksort. In particular, handling all the edges cases correctly made this algorithm very complicated.

\paragraph*{Acknowledgments} We thank \anonymize{Armin Weiss} for a useful hint in the 
two-pivot version. We also thank \anonymize{Timo Bingmann} who provided some source code used in the testing framework. Furthermore, we thank the anonymous reviewers for their suggestions that helped us in improving the presentation of this paper.


\begin{thebibliography}{10}
\providecommand{\url}[1]{\texttt{#1}}
\providecommand{\urlprefix}{URL }

\bibitem{Aumuller15}
Aum{\"{u}}ller, M.: On the Analysis of Two Fundamental Randomized Algorithms -
  Multi-Pivot Quicksort and Efficient Hash Functions. Ph.D. thesis, Technische
  Universit{\"{a}}t Ilmenau, Germany (2015),
  \url{http://www.db-thueringen.de/servlets/DocumentServlet?id=26263}

\bibitem{AumullerD16}
Aum{\"{u}}ller, M., Dietzfelbinger, M.: Optimal partitioning for dual-pivot
  quicksort. {ACM} Trans. Algorithms  12(2),  18:1--18:36 (2016),
  \url{http://doi.acm.org/10.1145/2743020}

\bibitem{AumullerDK16}
Aum{\"{u}}ller, M., Dietzfelbinger, M., Klaue, P.: How good is multi-pivot
  quicksort? {ACM} Trans. Algorithms  13(1),  8:1--8:47 (2016),
  \url{http://doi.acm.org/10.1145/2963102}

\bibitem{AxtmannWF017}
Axtmann, M., Witt, S., Ferizovic, D., Sanders, P.: In-place parallel super
  scalar samplesort (ipsssso). In: 25th Annual European Symposium on
  Algorithms, {ESA} 2017, September 4-6, 2017, Vienna, Austria. pp. 9:1--9:14
  (2017), \url{https://doi.org/10.4230/LIPIcs.ESA.2017.9}

\bibitem{Bentley86}
Bentley, J.L.: Programming pearls. Addison-Wesley (1986)

\bibitem{BrodalM05}
Brodal, G.S., Moruz, G.: Tradeoffs between branch mispredictions and
  comparisons for sorting algorithms. In: Proc. of the 9th International
  Workshop on Algorithms and Data Structures ({WADS}'05). pp. 385--395.
  Springer (2005), \url{http://dx.doi.org/10.1007/11534273_34}

\bibitem{EdelkampW16}
Edelkamp, S., Wei{\ss}, A.: Blockquicksort: Avoiding branch mispredictions in
  quicksort. In: 24th Annual European Symposium on Algorithms, {ESA} 2016,
  August 22-24, 2016, Aarhus, Denmark. pp. 38:1--38:16 (2016),
  \url{https://doi.org/10.4230/LIPIcs.ESA.2016.38}

\bibitem{ElmasryKS12}
Elmasry, A., Katajainen, J., Stenmark, M.: Branch mispredictions don't affect
  mergesort. In: Experimental Algorithms - 11th International Symposium, {SEA}
  2012, Bordeaux, France, June 7-9, 2012. Proceedings. pp. 160--171 (2012),
  \url{http://dx.doi.org/10.1007/978-3-642-30850-5\_15}

\bibitem{hennequin}
Hennequin, P.: Analyse en moyenne d'algorithmes: tri rapide et arbres de
  recherche. Ph.D. thesis, Ecole Politechnique, Palaiseau (1991)

\bibitem{HennessyP11}
Hennessy, J.L., Patterson, D.A.: Computer Architecture - {A} Quantitative
  Approach, 5th Edition. Morgan Kaufmann (2012)

\bibitem{hoare62}
Hoare, C.A.R.: Quicksort. Comput. J.  5(1),  10--15 (1962)

\bibitem{KaligosiS06}
Kaligosi, K., Sanders, P.: How branch mispredictions affect quicksort. In:
  Proc. of the 14th Annual European Symposium on Algorithms ({ESA}'06). pp.
  780--791. Springer (2006)

\bibitem{Kushagra14}
Kushagra, S., L{\'o}pez-Ortiz, A., Qiao, A., Munro, J.I.: Multi-pivot
  quicksort: Theory and experiments. In: Proc. of the 16th Meeting on
  Algorithms Engineering and Experiments ({ALENEX}'14). pp. 47--60. SIAM (2014)

\bibitem{MartinezR01}
Mart\'{\i}nez, C., Roura, S.: Optimal sampling strategies in quicksort and
  quickselect. SIAM J. Comput.  31(3),  683--705 (2001)

\bibitem{Musser97}
Musser, D.R.: Introspective sorting and selection algorithms. Softw., Pract.
  Exper.  27(8),  983--993 (1997)

\bibitem{NebelWM15}
Nebel, M.E., Wild, S., Martínez, C.: Analysis of pivot sampling in dual-pivot
  quicksort: A holistic analysis of yaroslavskiy’s partitioning scheme.
  Algorithmica pp. 1--52 (2015)

\bibitem{SandersW04}
Sanders, P., Winkel, S.: Super scalar sample sort. In: Proc. of the 12th Annual
  European Symposium on Algorithms ({ESA}'04). pp. 784--796. Springer (2004)

\bibitem{SedgewickAnalysis}
Sedgewick, R.: Implementing quicksort programs. Commun. ACM  21(10),  847--857
  (1978)

\bibitem{LomutoSwaps}
Wild, S.:
  \url{https://cs.stackexchange.com/questions/11458/quicksort-partitioning-hoare-vs-lomuto},
  accessed on August 8, 2018.

\bibitem{Wild16}
Wild, S.: Dual-pivot quicksort and beyond: Analysis of multiway partitioning
  and its practical potential. Ph.D. thesis, Technische Universität
  Kaiserslautern (2016)

\bibitem{Wild18}
Wild, S.: Dual-pivot and beyond: The potential of multiway partitioning in
  quicksort. it - Information Technology  60(3),  173--177 (2018),
  \url{https://doi.org/10.1515/itit-2018-0012}

\bibitem{Wild18b}
Wild, S.: Quicksort is optimal for many equal keys. In: Proceedings of the
  Fifteenth Workshop on Analytic Algorithmics and Combinatorics, {ANALCO} 2018,
  New Orleans, LA, USA, January 8-9, 2018. pp. 8--22 (2018),
  \url{https://doi.org/10.1137/1.9781611975062.2}

\bibitem{nebel12}
Wild, S., Nebel, M.E.: Average case analysis of java 7's dual pivot quicksort.
  In: Proc. of the 20th European Symposium on Algorithms ({ESA}'12). pp.
  825--836 (2012)

\end{thebibliography}

\appendix

\section{Running Time Plots for \texttt{i7}}
\label{sec:running:times:i7}

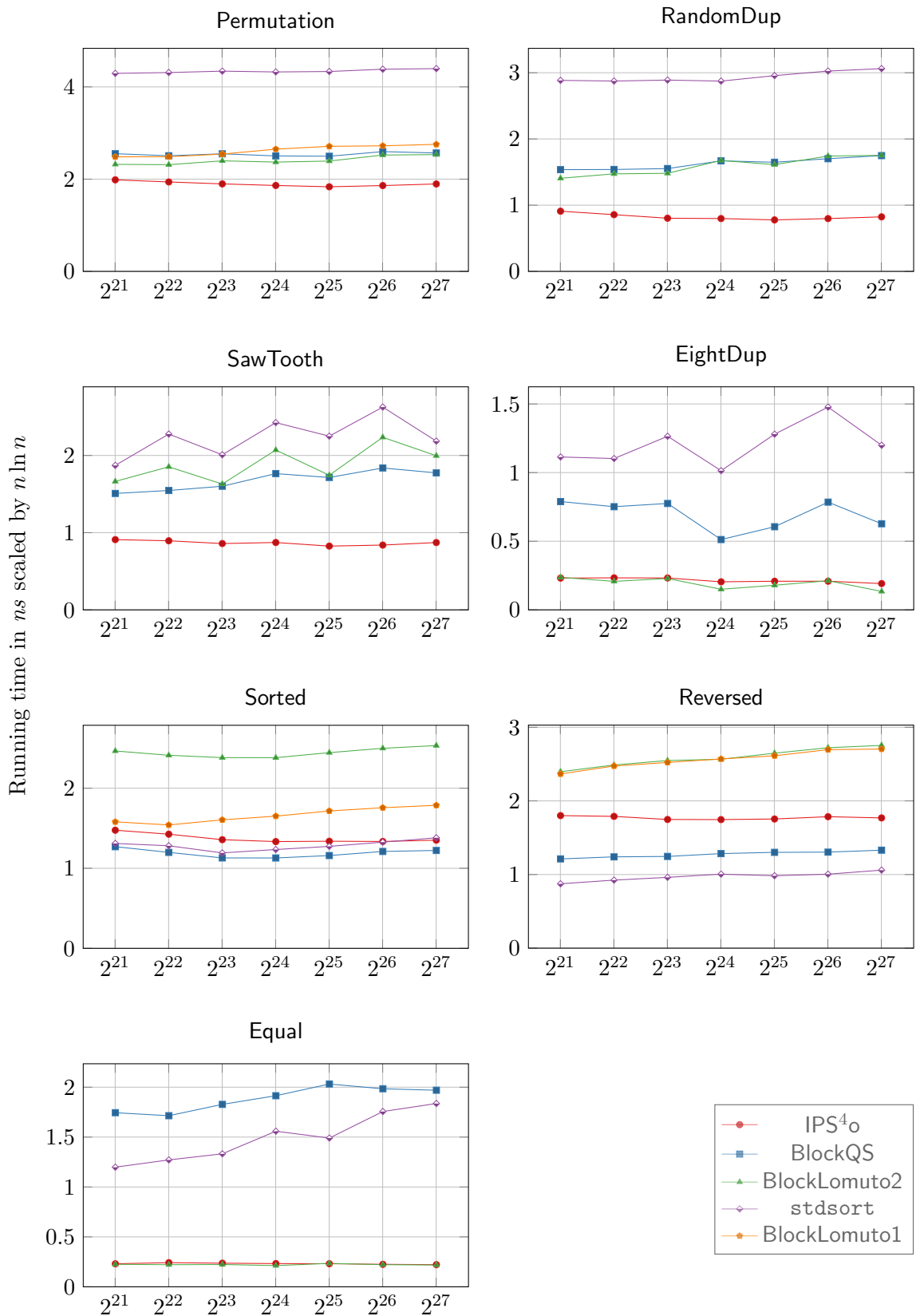
\begin{figure*}[h!]
\centering
\begin{tikzpicture}[every mark/.append style={mark size=1.5pt}]

\begin{groupplot}[group style = { group size = 2 by 4, group name = preciousss, vertical sep = 5 em},
                    height = 5.3cm,
                    width = 8cm,
                    xlabel = {},
                    ylabel = {},
                    xmode=log,
                    legend style = {opacity = 0.6, text opacity = 1},
                    ymin = 0,
                    log basis x = 2,
                    grid = both, grid style={line width=.1pt, draw=gray!30},
    major grid style={line width=.2pt,draw=gray!50},
                ]

\nextgroupplot[
                    title = {\textsf{Permutation}},
                ]
                
\addplot coordinates { (2097152.000000, 1.9848436) (4194304.000000, 1.9390262) (8388608.000000, 1.8970060) (16777216.000000, 1.8626975) (33554432.000000, 1.8342638) (67108864.000000, 1.8605298) (134217728.000000, 1.8955175) };
\addplot coordinates { (2097152.000000, 2.5515674) (4194304.000000, 2.5068817) (8388608.000000, 2.5494559) (16777216.000000, 2.5032971) (33554432.000000, 2.4965604) (67108864.000000, 2.5966693) (134217728.000000, 2.5672312) };
\addplot coordinates { (2097152.000000, 2.3207830) (4194304.000000, 2.3132448) (8388608.000000, 2.3967278) (16777216.000000, 2.3697239) (33554432.000000, 2.3917111) (67108864.000000, 2.5224191) (134217728.000000, 2.5322614) };
\addplot coordinates { (2097152.000000, 4.2935059) (4194304.000000, 4.3108233) (8388608.000000, 4.3413281) (16777216.000000, 4.3238409) (33554432.000000, 4.3334641) (67108864.000000, 4.3826967) (134217728.000000, 4.3959064) };
\addplot coordinates { (2097152.000000, 2.4831019) (4194304.000000, 2.4859310) (8388608.000000, 2.5432869) (16777216.000000, 2.6502347) (33554432.000000, 2.7112635) (67108864.000000, 2.7225143) (134217728.000000, 2.7541210) };
\nextgroupplot[
                    title = {\textsf{RandomDup}},
                ]
                
\addplot coordinates { (2097152.000000, .9092149) (4194304.000000, .8567863) (8388608.000000, .8030094) (16777216.000000, .7981607) (33554432.000000, .7775680) (67108864.000000, .7977060) (134217728.000000, .8237014) };
\addplot coordinates { (2097152.000000, 1.5367059) (4194304.000000, 1.5390102) (8388608.000000, 1.5527790) (16777216.000000, 1.6694858) (33554432.000000, 1.6482043) (67108864.000000, 1.7020164) (134217728.000000, 1.7485162) };
\addplot coordinates { (2097152.000000, 1.4064905) (4194304.000000, 1.4731096) (8388608.000000, 1.4812567) (16777216.000000, 1.6753081) (33554432.000000, 1.6124061) (67108864.000000, 1.7418701) (134217728.000000, 1.7528317) };
\addplot coordinates { (2097152.000000, 2.8848861) (4194304.000000, 2.8748464) (8388608.000000, 2.8894955) (16777216.000000, 2.8737764) (33554432.000000, 2.9563818) (67108864.000000, 3.0254973) (134217728.000000, 3.0623820) };
\nextgroupplot[
                    title = {\textsf{SawTooth}},
                ]
                
\addplot coordinates { (2097152.000000, .9098701) (4194304.000000, .8947788) (8388608.000000, .8591281) (16777216.000000, .8726328) (33554432.000000, .8259639) (67108864.000000, .8392753) (134217728.000000, .8719998) };
\addplot coordinates { (2097152.000000, 1.5077146) (4194304.000000, 1.5465149) (8388608.000000, 1.6010837) (16777216.000000, 1.7648107) (33554432.000000, 1.7145380) (67108864.000000, 1.8378579) (134217728.000000, 1.7743016) };
\addplot coordinates { (2097152.000000, 1.6638093) (4194304.000000, 1.8531131) (8388608.000000, 1.6294982) (16777216.000000, 2.0704382) (33554432.000000, 1.7438955) (67108864.000000, 2.2353528) (134217728.000000, 1.9960796) };
\addplot coordinates { (2097152.000000, 1.8713350) (4194304.000000, 2.2769721) (8388608.000000, 2.0087947) (16777216.000000, 2.4254749) (33554432.000000, 2.2500232) (67108864.000000, 2.6265224) (134217728.000000, 2.1853421) };
\nextgroupplot[
                    title = {\textsf{EightDup}},
                ]
                
\addplot coordinates { (2097152.000000, .2316033) (4194304.000000, .2331928) (8388608.000000, .2322140) (16777216.000000, .2043727) (33554432.000000, .2084774) (67108864.000000, .2086737) (134217728.000000, .1915397) };
\addplot coordinates { (2097152.000000, .7888271) (4194304.000000, .7516424) (8388608.000000, .7751183) (16777216.000000, .5121321) (33554432.000000, .6055251) (67108864.000000, .7846832) (134217728.000000, .6271316) };
\addplot coordinates { (2097152.000000, .2368447) (4194304.000000, .2075518) (8388608.000000, .2283631) (16777216.000000, .1500190) (33554432.000000, .1791887) (67108864.000000, .2122456) (134217728.000000, .1344052) };
\addplot coordinates { (2097152.000000, 1.1137924) (4194304.000000, 1.1023306) (8388608.000000, 1.2645585) (16777216.000000, 1.0138736) (33554432.000000, 1.2799465) (67108864.000000, 1.4772111) (134217728.000000, 1.1991769) };

\nextgroupplot[
                    title = {\textsf{Sorted}},
                ]
                
\addplot coordinates { (2097152.000000, 1.4756111) (4194304.000000, 1.4257362) (8388608.000000, 1.3572422) (16777216.000000, 1.3332955) (33554432.000000, 1.3376725) (67108864.000000, 1.3347259) (134217728.000000, 1.3510690) };
\addplot coordinates { (2097152.000000, 1.2698871) (4194304.000000, 1.1985627) (8388608.000000, 1.1295148) (16777216.000000, 1.1287436) (33554432.000000, 1.1589395) (67108864.000000, 1.2101915) (134217728.000000, 1.2227011) };
\addplot coordinates { (2097152.000000, 2.4652485) (4194304.000000, 2.4115876) (8388608.000000, 2.3819971) (16777216.000000, 2.3814402) (33554432.000000, 2.4444410) (67108864.000000, 2.4986723) (134217728.000000, 2.5328068) };
\addplot coordinates { (2097152.000000, 1.3101802) (4194304.000000, 1.2811926) (8388608.000000, 1.1921015) (16777216.000000, 1.2341726) (33554432.000000, 1.2735659) (67108864.000000, 1.3264451) (134217728.000000, 1.3809431) };
\addplot coordinates { (2097152.000000, 1.5797835) (4194304.000000, 1.5411991) (8388608.000000, 1.6050468) (16777216.000000, 1.6512126) (33554432.000000, 1.7156989) (67108864.000000, 1.7564762) (134217728.000000, 1.7866887) };
\nextgroupplot[
                    title = {\textsf{Reversed}},
                    legend style = { at = { (1, -0.7)}}
                ]
                
\addplot coordinates { (2097152.000000, 1.8002489) (4194304.000000, 1.7895577) (8388608.000000, 1.7476802) (16777216.000000, 1.7460718) (33554432.000000, 1.7540941) (67108864.000000, 1.7853660) (134217728.000000, 1.7687540) };
\addlegendentry{\ips};
\addplot coordinates { (2097152.000000, 1.2119044) (4194304.000000, 1.2400731) (8388608.000000, 1.2453787) (16777216.000000, 1.2841371) (33554432.000000, 1.3010917) (67108864.000000, 1.3042734) (134217728.000000, 1.3304928) };
\addlegendentry{\blockqs};
\addplot coordinates { (2097152.000000, 2.3962916) (4194304.000000, 2.4877290) (8388608.000000, 2.5480725) (16777216.000000, 2.5654437) (33554432.000000, 2.6480169) (67108864.000000, 2.7220471) (134217728.000000, 2.7537727) };
\addlegendentry{\lomutotwo};
\addplot coordinates { (2097152.000000, .8741632) (4194304.000000, .9240159) (8388608.000000, .9614578) (16777216.000000, 1.0053282) (33554432.000000, .9831560) (67108864.000000, 1.0050526) (134217728.000000, 1.0594450) };
\addlegendentry{stdsort};
\addplot coordinates { (2097152.000000, 2.3659899) (4194304.000000, 2.4738141) (8388608.000000, 2.5217891) (16777216.000000, 2.5696179) (33554432.000000, 2.6126744) (67108864.000000, 2.6959727) (134217728.000000, 2.7038201) };
\addlegendentry{\lomutoone};
\nextgroupplot[
                    title = {\textsf{Equal}},
                ]
                
\addplot coordinates { (2097152.000000, .2306205) (4194304.000000, .2417920) (8388608.000000, .2365135) (16777216.000000, .2324632) (33554432.000000, .2313597) (67108864.000000, .2252146) (134217728.000000, .2217422) };
\addplot coordinates { (2097152.000000, 1.7442317) (4194304.000000, 1.7141198) (8388608.000000, 1.8278765) (16777216.000000, 1.9142923) (33554432.000000, 2.0315279) (67108864.000000, 1.9840101) (134217728.000000, 1.9694681) };
\addplot coordinates { (2097152.000000, .2252154) (4194304.000000, .2213104) (8388608.000000, .2234653) (16777216.000000, .2121657) (33554432.000000, .2345155) (67108864.000000, .2216178) (134217728.000000, .2182209) };
\addplot coordinates { (2097152.000000, 1.1984734) (4194304.000000, 1.2713427) (8388608.000000, 1.3323795) (16777216.000000, 1.5587361) (33554432.000000, 1.4901523) (67108864.000000, 1.7560462) (134217728.000000, 1.8364343) };
\end{groupplot}
    \path (top-|current bounding box.west)-- node[anchor=south,rotate=90] {Running time in \emph{ns} scaled by $n \ln n$} (bot-|current bounding box.west);
\end{tikzpicture}
\caption{Running time plots on \texttt{i7}. The $x$-axis represents the number of items, the $y$-axis shows the running time in nanoseconds scaled by $n \ln n$.}
\label{plot:running:times:i7}
\end{figure*}

The machine \texttt{i7}  was equipped with a 4-core Intel Core i7-4790 clocked at 3.6 GHz, with 8 MB L3 Cache and  32GB RAM. \textsf{i7} was running CentOS 7 with Linux kernel 3.10 and code was compiled using \texttt{gcc} version 5.1.1. Figure~\ref{plot:running:times:i7} presents an overview over the running time measurements that we obtained. While \texttt{i7} turns out to be faster than \texttt{Xeon}, differences in running time are comparable to Section~\ref{sec:evaluation}.

\end{document}